\documentclass[journal]{IEEEtran}                                        

\usepackage{graphics} % for pdf, bitmapped graphics files
\usepackage{epsfig} % for postscript graphics files
\usepackage{mathptmx} % assumes new font selection scheme installed
\usepackage{times} % assumes new font selection scheme installed
\usepackage{amsmath} % assumes amsmath package installed
\allowdisplaybreaks[0]
\usepackage{amssymb}  % assumes amsmath package installed
\usepackage{soul, color} % To highlight texts with the command `hl'
\usepackage{algorithm}
\usepackage{algorithmic}

\usepackage{amsfonts,amsmath,amssymb,amsthm}
\usepackage{bm,nicefrac}

\usepackage{upquote}
\usepackage{mathrsfs}
\usepackage{lmodern}
\usepackage{bm}

\usepackage{adjustbox}
\usepackage{multirow}
\usepackage{adjustbox}
\usepackage{threeparttable}
\usepackage{mathtools}
\allowdisplaybreaks
\usepackage{balance}
\usepackage {tikz}
\usetikzlibrary {positioning}
\usepackage{booktabs}
\usepackage[scale=2]{ccicons}
\usepackage{animate}
\usepackage{graphicx}
\usepackage{media9}
\usepackage{pgfplots}

\usepackage{array}
\usepackage{xcolor}
\usepgfplotslibrary{dateplot}

\usepackage{xspace}

\usepackage[numbers,sort&compress]{natbib}

\usepackage{etoolbox}

\usepackage{mathrsfs}
\usepackage{pgfplots}
\usepgfplotslibrary{dateplot}
\usepackage{xspace}
\usepackage{mathptmx}

\usepackage{bbold}

\usepackage{mathtools}
\usepackage{amsmath,amsthm,amssymb,hyperref}
\newtheorem{assumption}{Assumption}
\newtheorem{theorem}{Theorem}%[section]

\newtheorem{lemma}[theorem]{Lemma}

\title{\LARGE \bf
Identifying the Influential Inputs for Network Output Variance Using Sparse Polynomial Chaos Expansion}

\author{Zhanlin Liu$^{1}$, Ashis G. Banerjee$^{2}$, and Youngjun Choe$^{1}$% <-this % stops a space
\thanks{$^{1}$Z. Liu and Y. Choe (e-mail: ychoe@uw.edu) are with the Department of Industrial \& Systems Engineering, University of Washington, Seattle WA 98195, USA.}%
\thanks{$^{2}$A. G. Banerjee is with the Department of Industrial \& Systems Engineering and the Department of Mechanical Engineering, University of Washington, Seattle WA 98195, USA.}%
}

\IEEEoverridecommandlockouts

\usepackage{tikz}
\usepackage{textcomp}
\usepackage{hyperref}
\usepackage{lipsum}

\newcommand\copyrighttext{%
  \footnotesize \textcopyright 2020 IEEE. Personal use of this material is permitted.  Permission from IEEE must be obtained for all other uses, in any current or future media, including reprinting/republishing this material for advertising or promotional purposes, creating new collective works, for resale or redistribution to servers or lists, or reuse of any copyrighted component of this work in other works.}
\newcommand\copyrightnotice{%
\begin{tikzpicture}[remember picture,overlay]
\node[anchor=south,yshift=10pt] at (current page.south) {\fbox{\parbox{\dimexpr\textwidth-\fboxsep-\fboxrule\relax}{\copyrighttext}}};
\end{tikzpicture}%
}
\pgfplotsset{compat=1.14}

\usepackage{nomencl}
\begin{document}

%\IEEEoverridecommandlockouts

\maketitle
\copyrightnotice

%\thispagestyle{plain}
%\pagestyle{plain}

%\IEEEpeerreviewmaketitle

%%%%%%%%%%%%%%%%%%%%%%%%%%%%%%%%%%%%%%%%%%%%%%%%%%%%%%%%%%%%%%%%%%%%%%%%%%%%%%%%%up to 200 words for Regular Papers %https://www.ieee-ras.org/publications/t-ase/information-for-authors-t-ase
\begin{abstract}
Sensitivity analysis (SA) is an important aspect of process automation. It often aims to identify the process inputs that influence the process output's variance significantly. Existing SA approaches typically consider the input-output relationship as a black-box and conduct extensive random sampling from the actual process or its high-fidelity simulation model to identify the influential inputs. In this paper, an alternate, novel approach is proposed using a sparse polynomial chaos expansion-based model for a class of input-output relationships represented as directed acyclic networks. The model exploits the relationship structure by recursively relating a network node to its direct predecessors to trace the output variance back to the inputs. It, thereby, estimates the Sobol indices, which measure the influence of each input on the output variance, accurately and efficiently. Theoretical analysis establishes the validity of the model as the prediction of the network output converges in probability to the true output under certain regularity conditions. Empirical evaluation on two manufacturing {\color{black}processes and a flooding process} shows that the model estimates the Sobol indices accurately with far fewer observations than state-of-the-art {\color{black}black-box} methods.

\textit{Note to Practitioners--}%https://www.ieee-ras.org/publications/t-ase/information-for-authors-t-ase/50-publications/t-ase/236-t-ase-note-to-practitioners
This paper is motivated by the problem of automated identification of the inputs that influence the variance of the output for networked processes without feedback control. Such processes arise in various natural and engineered systems, of which manufacturing operations {\color{black}and flood mitigation} are of particular interest to us, where the output variance represents the uncertainty in productivity, quality{\color{black}, or cost}. Therefore, influential inputs identification allows us to quantify the effects of the various process parameters, such as operating conditions {\color{black}and physical} properties, in determining the uncertainties in the process outputs. We show that our identification method is guaranteed to quantify the effects accurately, and is expected to do so more efficiently (with fewer experimental observations) than widely used stochastic sampling technique{\color{black}s}. In the future, we will like to evaluate the usefulness of the developed method on large-scale manufacturing and supply chain networks in actual production facilities {\color{black}as well as on critical infrastructures subject to cascading failures.}

\end{abstract}

{
\color{black}
%\mbox{}

% \nomenclature{$\mathbb{N}$}{natural number set;}
% \nomenclature{$\mathbb{R}$}{rational number set;}
\nomenclature[01]{$\boldsymbol{X}_{j}$}{input vector of the $j^{th}$ observation, $j=1,\ldots,m$}
\nomenclature[02]{$Y_{j}$}{output of the $j^{th}$ observation, $j=1,\ldots,m$}
\nomenclature[03]{$S_{x_{j}}$}{first-order Sobol index of the output $y$ with respect to the input $x_{j}$}
\nomenclature[04]{$ST_{x_{j}}$}{total Sobol index of the output $y$ with respect to the input $x_{j}$}
\nomenclature[05]{$G$}{directed acyclic graph representing the network-structured process of interest}
\nomenclature[06]{$E$}{collection of all the directed edges in $G$}
\nomenclature[07]{$V$}{collection of all the nodes in $G$}
\nomenclature[08]{$S(G)$}{source nodes in $G$}
\nomenclature[111]{$\boldsymbol{x}_{S(G)}$}{network inputs}
\nomenclature[10]{$x_{v_{i}}$}{random variable represented by the node $
v_{i}\in V$}
\nomenclature[11]{$\boldsymbol{x}_{\boldsymbol{v}}$}{$\left(x_{v_{i}}\right)_{v_{i}\in \boldsymbol{v}}$ for $\boldsymbol{v} \subseteq V$}
\nomenclature[12]{$\boldsymbol{\xi}$}{network inputs that influence the output $y$}
%\nomenclature{$P+1$}{number of orthonormal polynomials in a PCE;}
%\nomenclature{$p$}{pre-specified highest polynomial order in a PCE;}
%\nomenclature{$\theta_{i}$}{the $i^{th}$ PCE coefficient in a PCE;}
%\nomenclature[y]{$\hat{y}^{(l)}$}{estimated output after the $l^{th}$ iteration;}
\nomenclature[13]{$\mathscr{P}(\boldsymbol{x}_{\boldsymbol{v}})$}{$\boldsymbol{x}_{ \left\{v_{i} \in V: v_{j}\in \boldsymbol{v}, \langle v_{i}, v_{j} \rangle \in E \right \} \cup \left(\boldsymbol{v} \cap S(G) \right)}$}
\nomenclature[14]{$\mathscr{P}^{l}(\boldsymbol{x}_{\boldsymbol{v}})$}{$\mathscr{P}(\cdot)$ applied on $\boldsymbol{x}_{\boldsymbol{v}}$ $l\in \mathbb{N}$ times}
\nomenclature[15]{$L$}{$\inf\{l\in\mathbb{N}: \mathscr{P}^{l}(y) = \mathscr{P}^{l+1}(y) \}$ for the output $y$}%denotes the total number of iterations.  

}

%\vspace{-0.3cm}
\vspace{-0.2cm}
%%%%%%%%%%%%%%%%%%%%%%%%%%%%%%%%%%%%%%%%%%%%%%%%%%%%%%%%%%%%%%%%%%%%%%%%%%%%%%%%
\section{INTRODUCTION}\label{sec:introduction}

\IEEEPARstart{U}{ncertainty} quantification plays a critical role in controlling the uncertainties in process automation \cite{djurdjanovic2006stream,marvel2013performance}. Automated processes that neglect important uncertainties are, at minimum, unstable, and, in the worst case, have catastrophic process outcomes in terms of both quality and productivity. To quantify how such uncertainties propagate through the process, sensitivity analysis is widely used in process automation. 
%As an aspect of UQ, sensitivity analysis (SA) studies how the uncertainty of a model output is propagated from model inputs. 
The sensitivity analysis of this study focuses on characterizing how the process output's variance is propagated from the inputs. This type of analysis is ubiquitous in different areas of automation science and engineering, such as thermodynamics {\color{black}\cite{avdonin2018uncertainty}}, electromagnetism \cite{chen2015sensitivity}, power systems \cite{ni2018variance}, building systems \cite{li2017sensitivity}, and manufacturing \cite{huang2010process}. 

How sensitive a random variable (e.g., process output) is with respect to another random variable (e.g., process input) is measured using a \textit{sensitivity index}. Arguably, the most widely used sensitivity indices are 
the \textit{Sobol indices} %, which is proposed based on the Sobol decomposition in 
\cite{sobol:1993}, which quantify how the independent inputs apportion the variance of the output. In practice, simple random sampling (from the actual process) or Monte Carlo sampling (from the simulation model of the process) is most commonly used to estimate the Sobol indices, where many observations of the inputs and output are available. A more resource-efficient alternative is to construct a model (or, metamodel) of the actual process (or, its simulation model when the computational cost is expensive), and use the model (or, metamodel) to estimate the Sobol indices \cite{sudret2012meta}. 

Among such models, \textit{polynomial chaos expansion} (PCE) is particularly conducive to estimating the Sobol indices, as detailed in Sec.~\ref{sec:background}. In essence, the PCE expands a random variable (e.g., process output) in terms of orthonormal polynomials in other random variables (e.g., process inputs), and the expansion's coefficients directly yield convergent estimators of the Sobol indices thanks to the orthogonality of the polynomials in a Hilbert space. 

However, such models, including PCE, typically consider the input-output relationship of a process as a black-box for sensitivity analysis. This approach, while generally applicable to many processes, misses the opportunity to leverage the scientific/engineering knowledge of how the process actually works. Opening the black-box and utilizing the knowledge of the inner working can enable more effective modeling of the process, leading to more accurate and efficient sensitivity analysis. 

To this end, this study considers a broad class of processes whose input-output relationships are expressed as \textit{directed acyclic graphs} (DAGs), also known as directed acyclic networks. Since network-structured processes are ubiquitous in practice, several real-world systems can potentially benefit from this study, such as biological networks \cite{prill2005dynamic}, supply chain systems \cite{mogale2016two}, manufacturing operations \cite{he2016optimal}, and process industries, in general \cite{agrawal2014energy}. In particular, this study uses two manufacturing processes (welding and injection molding) in Sec.~\ref{sec:appli} and a flooding process in the article's online supplementary document for illustration purposes.
%In practice, causal relationships are present in a network structure (often, directed acyclic graph) in a variety of science and engineering applications, such as , , , and . Identifying influential inputs in explaining output variance of a network help control the variability and improve the productivity or quality. 

%However, the existing PCEs typically take the input-output relationships as a black-box and construct PCEs with respect to all the possible inputs. This increases the requirement on the number of observations of the inputs and output since it might involve unnecessary terms which are not present in the practical network structures. 

Consequently, the main contribution of this paper is in the development of a novel model, called sparse network PCE (SN-PCE), 
%which effectively models a process represented as a DAG 
to accurately estimate the Sobol indices of the process output (sink node in the DAG) with respect to the process inputs (source nodes in the DAG). {\color{black}To the best of our knowledge, SN-PCE provides the most efficient way to estimate the Sobol indices for any process represented as a DAG.} The estimated Sobol indices allow us to identify the inputs that significantly influence the output variance. The proposed model is validated through a theoretical analysis, which establishes that the prediction of the output converges in probability to the true output under certain regularity conditions (without imposing impractical restrictions such as parametric relationships among the network variables). The model is also empirically validated through sensitivity analysis of {\color{black}three real-world} processes, where it is shown to accurately estimate the Sobol indices with much fewer observations of the process inputs and output as compared to state-of-the-art {\color{black}black-box methods}.  
%to identify influential independent inputs for the output variance in a engineering system which can be formed as a directed acyclic graph (DAG). We propose network sensitivity indices to quantify the sensitivity indices for the influential inputs. We first review the definitions on directed acyclic graphs. Then we study the feasibility and limitation of using an existing PCE to address the problem. After that, we propose a network PCE (nPCE), which leverages the knowledge about the network structure, to identify the influential inputs. With the proposed nPCE, it significantly reduces the number of the minimally required observations. In addition, we further propose a sparse network PCE (SN-PCE) model to improve the efficiency of the nPCE. The SN-PCE also keeps the principle of parsimony on approximating the network output with respect to the inputs. Theoretical results are provided to validate the use of the model for identifying the inputs that most influence the output variance through the network. The model is empirically evaluated in the analysis of two different manufacturing processes, namely, the welding process and the injection molding process.    

The remainder of this paper is organized as follows. Sec.~\ref{sec:background} briefly reviews the technical background on the PCE and Sobol indices.  Sec.~\ref{sec:methodology} starts with formal definitions regrading the DAG and presents the following three PCEs to model a process represented as a DAG: na\"{\i}ve PCE, network PCE, and SN-PCE. {\color{black}Sec.~\ref{sec:methodology} also} discusses the theoretical properties of the PCEs. In Sec.~\ref{sec:appli}, the PCEs are empirically evaluated with two manufacturing applications. %In the end, we provide a few concluding remarks and a discussion of future research directions in 
Sec.~\ref{sec:conclusion} concludes the paper with a discussion on future research directions.

\vspace{-0.3cm}
\section{Technical Background}\label{sec:background}
In this section, we first formally define process inputs and output that bear uncertainties. Then, we present the PCE and sparse PCE. We review the construction of orthonormal polynomials for the PCE, the Hoeffding decomposition, and the Sobol indices. We also review how to estimate the Sobol indices using the PCE. %for models with independent inputs. 

\subsection{Input random vector and output random variable}
In this paper, we generally use the same notations as \cite{rahman2018polynomial}, including $\mathbb{N}_{0}:=\mathbb{N}\cup \{0\}$. $\mathbb{A}^{n}\subseteq \mathbb{R}^{n}$, $n\in \mathbb{N}$, denotes a bounded or unbounded subdomain of $\mathbb{R}^{n}$. Let $\Omega$ be a sample space, and $\mathcal{F}$ be a $\sigma$-algebra on $\Omega$. $(\Omega, \mathcal{F}, \mathcal{P})$ is a probability space with a probability measure $\mathcal{P}: \mathcal{F} \rightarrow [0,1]$. The Borel $\sigma$-algebra on $\mathbb{A}^{n} \subseteq \mathbb{R}^{n}$ is represented as $\mathcal{B}^{n} \coloneqq \mathcal{B}(\mathbb{A}^{n})$. An $\mathbb{A}^{n}$-valued input random vector $\boldsymbol{x}\coloneqq(x_{1}, \ldots, x_{n}):(\Omega , \mathcal{F}) \rightarrow (\mathbb{A}^{n}, \mathcal{B}^{n})$ describes the uncertainties of $n$ process inputs of interest. 

To model a process output $y$ using the PCE in $\boldsymbol{x}$, we assume that $y$ is a function of $\boldsymbol{x}$ and is a square-integrable random variable defined on the same probability space $(\Omega, \mathcal{F}, \mathcal{P})$, written $y\!\left(\boldsymbol{x}\right) \in \mathcal{L}^{2}\!\left(\Omega, \mathcal{F}, \mathcal{P}\right)$. 

\subsection{Polynomial chaos expansion (PCE)}\label{subsec:pce}
PCE approximates $y$ using a finite number of orthonormal polynomials in $\boldsymbol x$ as follows:
\begin{equation}
\label{eq:exp}
\hat{y}= f(\boldsymbol x) \approx \sum_{i=0}^{P}\theta_{i}\psi_{i}(\boldsymbol x), 
\end{equation}
where $\theta_{i}$ and $\psi_{i}(\boldsymbol x)$, $i= 0,1,2,\ldots,P$, denote the PCE coefficients and orthonormal polynomials in $\boldsymbol x$, respectively. $P+1$ is the number of the orthonormal polynomials %, which truncates at the $(P+1)^{th}$ term, where 
and equals $\binom{n+p}{n}$ for the pre-specified highest polynomial order $p$ and the dimension of $\boldsymbol x$, $\dim(\boldsymbol x)=n$. Thus, $P+1$ increases exponentially in $n$ and $p$. As $P$ increases, the approximation error in eq.~\eqref{eq:exp} tends to zero \cite{Cameron:1947}.

In this paper, we adopt a regression-based method to obtain the PCE coefficients by solving an overdetermined linear system of equations in the least-squares sense as follows \cite{pettersson2015polynomial}: 
\begin{equation}
\begin{aligned}
\label{eq:over_lin}
 & \underset{\boldsymbol{\theta}\in\mathbb{R}^{P+1}}{\mathrm{argmin}} \sum_{j=1}^{m}\left(Y_{j} - 
 \sum_{i=0}^{P}\theta_{i}\psi_{i}\left(\boldsymbol{X}_{j}\right)\right)^{2}, \\ 
\end{aligned}
\end{equation}
where $\boldsymbol{\theta}$ denotes $\left(\theta_{0}, \theta_{1},\ldots, \theta_{P}\right)$. $Y_{j}$ and $\boldsymbol{X}_{j}$ represent the output and the input vector of the $j^{th}$ observation, $j=1,\ldots,m$, respectively.

Note that in contrast to typical regression models whose coefficients explain how the \textit{expectation} of the output would conditionally change when the inputs are varied, the PCE coefficients are used to interpret how the  \textit{variance} of the output is apportioned to the inputs. 

Thanks to the orthogonality of the orthonormal polynomials, the lower order moments of the output $y$ are approximated using the PCE coefficients in eq.~\eqref{eq:exp} as follows: 
\begin{equation}
\begin{aligned}
\label{eq:mom}
&\mathbb{E}(y) \approx \theta_{0},\\
&Var(y) \approx \sum_{i=1}^{P}\theta_{i}^2,
\end{aligned}
\end{equation}
where $\mathbb{E}$ is the expectation operator with respect to the probability measure $\mathcal{P}$. The approximation errors in eq.~\eqref{eq:mom} tend to zero as $P$ in the PCE in eq.~\eqref{eq:exp} increases. 

\subsection{Sparse PCE}

The sparse PCE is extensively studied in the recent literature {\color{black} \cite{blatman2010efficient, blatman2011adaptive, jakeman2015enhancing, peng2016polynomial,shao2017bayesian, cheng2018sparse}}. In this paper, we use $l_{1}$ minimization, specifically the least absolute shrinkage and selection operator (LASSO), 
to obtain a more parsimonious model than the model from eq.~\eqref{eq:over_lin}, by solving the following problem:
\begin{equation}
\begin{aligned}
\label{eq:lasso}
&\underset{\boldsymbol{\theta}\in\mathbb{R}^{P+1}}{\mathrm{argmin}}\, \sum_{i=0}^{P}\left|\theta_{i}\right|,\\
& \text{such that } \frac{\sum_{j=1}^{m}\left(Y_{j} - 
 \sum_{i=0}^{P}\theta_{i}\psi_{i}\left(\boldsymbol{X}_{j}\right)\right)^{2}} {\sum_{j=1}^{m}\left(Y_{j} - 
 \bar{Y}\right)^{2}} \leq \Gamma, 
\end{aligned} 
\end{equation} 
where $\bar{Y} = \sum_{j=1}^{m}Y_{j}/m$. The parameter $\Gamma$ constrains the goodness-of-fit for the sparse PCE (e.g., $\Gamma=0$ requires a perfectly fitting model and $\Gamma = 1$ allows the model to be as simple as a constant model). The two application examples in Sec.~\ref{sec:appli} use $\Gamma$ of 0.001.

\subsection{Construction of multivariate orthonormal polynomials}

A variety of PCEs have been proposed to construct orthonormal polynomials considering different types of input distributions.  
The Wiener chaos expansion, known as the first PCE, uses Hermite polynomials for independent Gaussian-distributed inputs \cite{Wiener:1938}. Later PCEs allow for different input distributions and include %Considering the complexity of natural phenomena, where the normality assumption may not necessarily hold, different PCEs have been explored, including 
the generalized PCE (gPCE) \cite{Xiu:2002}, the multi-element gPCE (ME-gPCE) \cite{Xiaoliang:2006}, the moment-based arbitrary PCE (aPCE) \cite{Oladyshkin:2012}, %transformation-based polynomial chaos expansion \cite{feinberg2018multivariate}, 
and the Gram-Schmidt based PCE (GS-PCE) \cite{Witteveen:2007}. 
In particular, GS-PCE, which accounts for dependent inputs following arbitrary distributions \cite{Navarro:2014}, provides the basis for constructing the proposed model in this paper. 

In this work, process inputs are assumed to be mutually independent. However, the other variables in the process (represented as a network) are allowed to be dependent on others. When the variables in $\boldsymbol{x}$ are \textit{mutually independent}, the orthonormal polynomials are directly constructed as the tensor products of the univariate orthonormal polynomials as follows: 
\begin{equation}
\label{eq:multi}
\psi_{i} (\boldsymbol x)  = \psi_{ \boldsymbol{\alpha}_{i} } (\boldsymbol x)  := \prod_{j=1}^{n}\psi_{\alpha_{ij}}\left(x_{j}\right), 
\end{equation}
where $\boldsymbol{\alpha}_{i} := (\alpha_{i1}, \alpha_{i2}, \ldots, \alpha_{in})$. $\psi_{\alpha_{ij}}\left(x_{j}\right)$ represents the $\alpha_{ij}$-th order orthonormal polynomial in input $x_{j}$. $\boldsymbol{\alpha}_{i}$ is the $i$-th arbitrary vector satisfying $\left|\boldsymbol{\alpha}_{i}\right|:=\sum_{j=1}^{n} \alpha_{ij} \leq p$, where $p$ is the highest order of the polynomials in the PCE.
%In this case, we only need to construct the orthonormal polynomials for each univariate input $x_{j}, j =1,2,\ldots,n$. %For each $e_{j}(\xi_{i}) = \xi_{i}^{j}$, where $j=1,2,\ldots,p$. 
When variables are \textit{dependent on each other}, we construct orthonormal polynomials using the modified Gram-Schmidt algorithm, as proposed in \cite{liu2018data}. 

\subsection{Hoeffding decomposition and Sobol indices}\label{subsec:Sobol_ind}
Suppose the inputs in the $n$-dimensional vector $\boldsymbol{x}$ are mutually independent and $y=f(\boldsymbol x) \in \mathcal{L}^{2}\!\left(\Omega, \mathcal{F}, \mathcal{P}\right)$. %is a square-integrable function of $\boldsymbol{x}$. 
Denote the probability density function of $\boldsymbol x$ as $\mu(\boldsymbol x)$. The Hoeffding decomposition of %$f\in L^{2}(\Omega_{n},\mu(\boldsymbol \xi))$ 
$f(\boldsymbol{x})$ is then defined as follows \cite{sobol:1993,chastaing2012generalized}: 
\begin{equation}
\begin{aligned}
\label{eq:anova}  
f(\boldsymbol x) := \sum_{u \subseteq \{1,2,\ldots, n\}}f_{u}(\boldsymbol x_{u}), 
\end{aligned}
\end{equation}
%%%
where $f_{\emptyset} := f_{0}$ is a constant and $\boldsymbol x_{u}:=\left(x_j\right)_{j\in u}$ for $u\neq \emptyset$. Such decomposition is unique if and only if the summands in eq.~\eqref{eq:anova} are orthogonal to each other as follows \cite{sobol:1993}: % check citation
\begin{equation}
\begin{aligned}
\label{eq:sob21}  
\int f_{u}(\boldsymbol x)f_{v}(\boldsymbol x) \mu(\boldsymbol x) d\boldsymbol x = 0, \forall u, v \subseteq \{1,2,\ldots, n\}, v\neq u.
\end{aligned}
\end{equation}
Due to the orthogonal property in eq.~\eqref{eq:sob21}, the functional decomposition of $Var(y)$ is expressed as follows \cite{crestaux:2009}: 
\begin{equation}
\begin{aligned}
\label{eq:anova2}
Var(y) &= \int f^{2}(\boldsymbol x)\mu(\boldsymbol x)d \boldsymbol x - f_{0}^{2}\\
&= \sum_{\substack{ u \subseteq \{1,2,\ldots, n\} \\ u\neq \emptyset }}D_{u}(y), \\
\end{aligned}\nonumber
\end{equation}
%%%
where 
\begin{equation}
\begin{aligned}
\label{eq:anova3}
D_{u}(y) &:=  \int f^{2}_{u}(\boldsymbol x_{u})\mu(\boldsymbol x_{u}) d\boldsymbol x_{u} \\
&=Var\left(\mathbb{E}\left(y|\boldsymbol x_{u}\right)\right) - \sum_{\substack{ v \subset u\\ v \neq u \\ v \neq \emptyset }}D_{v}(y). 
\end{aligned}\nonumber
\end{equation}

Based on the decomposition, our sensitivity analysis considers the \textit{first-order} Sobol index $S_{x_{j}}$ and \textit{total} Sobol index $ST_{x_{j}}$ of $y$ with respect to $x_{j}$ defined as follows: 
\begin{equation}
\begin{aligned}
\label{eq:sobtot}
S_{x_{j}} &:= \frac{D_{\{j\}}(y)}{Var(y)},\\
ST_{x_{j}} &:= \sum_{u \ni x_{j}}S_{u}, \\
\end{aligned}\nonumber
\end{equation}
where $S_{u} := {D_{u}(y)}/{Var(y)}.$ The first-order Sobol index $S_{x_{j}}$ measures the \textit{main effect} of input $x_{j}$ on the output variance $Var(y)$. The total Sobol index $ST_{x_{j}}$ measures the \textit{total contribution} of $x_{j}$ to $Var(y)$ including its main effect \textit{and} interactions with other inputs \cite{homma1996importance}. 

\subsection{PCE-based Sobol indices}\label{subsec:PCE_Sobol}
The Sobol indices can be estimated directly using the PCE coefficients in eq.~\eqref{eq:exp}, making PCE particularly useful for the sensitivity analysis \cite{Sudret:2008}. The first-order Sobol index is estimated as follows: 
\begin{equation}
\begin{aligned}
\label{eq:sobeqs}
S_{x_{j}} &\approx \frac{\sum_{\boldsymbol{\alpha}_{i} \in \mathscr{A}_{\{j\}}}\theta_{\boldsymbol \alpha_{i}}^{2}}{\sum_{i=1}^{P}\theta_{i}^{2}},\\ 
\end{aligned}
\end{equation}
where $\theta_{\boldsymbol{\alpha}_{i}}$ is the PCE coefficient with respect to $\psi_{\boldsymbol{\alpha}_{i}}(\boldsymbol{x})$ in eq.~\eqref{eq:multi} and 
\begin{equation}
\begin{aligned}
%\label{eq:Au1}
\begin{multlined}[0.45\linewidth]
\mathscr{A}_{u} :=  \left\{\boldsymbol \alpha_{i} \in \mathbb{N}^{n}: \alpha_{ij}\neq 0 \leftrightarrow j \in u, |\boldsymbol \alpha_{i}| \leq p\right\}.
\end{multlined} \nonumber \\
\end{aligned}
\end{equation}
The total Sobol index is estimated as follows:
\begin{equation}
\begin{aligned}
\label{eq:sobeqs2}
ST_{x_{j}} &\approx \sum_{\mathscr{A}_{u \ni j} } S_{\mathscr{A}_{u}},
\end{aligned}
\end{equation}
where 
\begin{equation}
\begin{aligned}
\label{eq:sobeqs3}
S_{\mathscr{A}_{u}} &\approx \frac{\sum_{\boldsymbol{\alpha}_{i} \in \mathscr{A}_{u}}\theta_{\boldsymbol \alpha_{i}}^{2}}{\sum_{i=1}^{P}\theta_{i}^{2}}.
\end{aligned}\nonumber
\end{equation}

\section{Methodology}
\label{sec:methodology}
In this section, we first define notations on the directed acyclic graph (DAG), which represents the network-structured process of interest. Then, we present three models to estimate the Sobol indices for the process output with respect to the process inputs. The first model called na\"{\i}ve PCE is a baseline model and directly approximates the process output as a function of the process inputs, viewing the process as a black-box. The second model called network PCE uses the network structure of the process to effectively approximate the process output in terms of the process inputs. The third model called sparse network PCE (SN-PCE) imposes sparsity on the second model to use even fewer observations for the sensitivity analysis than the other models. 

{\color{black} In addition,} we show that predicted outputs from na\"{\i}ve PCE and network PCE converge to the true network output in probability under certain regularity conditions. This validates the use of the PCEs for estimating Sobol indices to conduct a sensitivity analysis. Because SN-PCE is a sparsity-imposed version of network PCE, its validity follows from the validity of network PCE and the sprase PCE.

\subsection{Directed acyclic graph}
Let $G = DAG(V,E)$ be the directed acyclic graph that represents the network-structured process of interest. $V= \{v_{1}, v_{2}, \ldots, v_{\left|V\right|}\}$ is the collection of all the nodes in $G$, where $\left|V\right|$ denotes the number of the nodes. Let $x_{v_{i}}$ denote the random variable represented by the node $
v_{i}\in V$ and, for $\boldsymbol{v} \subseteq V$, $\boldsymbol{x}_{\boldsymbol{v}} :=\left(x_{v_{i}}\right)_{v_{i}\in \boldsymbol{v}}$.  $E \subseteq V \times V $ is the collection of all the directed edges in $G$, encoding all dependencies between the nodes.
For example, $\langle v_{i}, v_{j} \rangle \in E $ implies that there is an edge from node $v_{i}$ to node $v_{j}$, and hence $x_{v_{j}}$ depends on $x_{v_{i}}$. The adjacency matrix $\boldsymbol{A}$ is defined as follows:
\begin{equation}
\begin{aligned}
    A_{ij} := \begin{cases} 1 & \langle v_{i}, v_{j}\rangle \in E \\
    0 & \langle v_{i}, v_{j}\rangle \notin E.
    \end{cases}
\end{aligned} \nonumber
\end{equation}
If $A_{ij} =1$, then $v_{i}$ is called a \textit{direct predecessor} of $v_{j}$. If $\sum_{j=1}^{\left|V\right|}A_{ji}=0$, 
$v_{i}$ is called a \textit{source node}. %On the other hand, if $k_{out}(v_{i}) = 0$, then 
If $\sum_{j=1}^{\left|V\right|}A_{ij}=0$, $v_{i}$ is termed as a \textit{sink node}. Let $S(G)$ denote all the source nodes in $G$. We call the variables in $\boldsymbol{x}_{S(G)}$ \textit{network inputs} and the variables represented by the sink nodes \textit{network outputs}. The node corresponding to the network output $y$ is denoted by $v_y$.

We say that there exists a \textit{path} from $v_i$ to $v_j$ if and only if either $A_{ij} = 1$ or there exists a sequence of nodes $\left(v_{k_1}, \ldots, v_{k_\tau}\right)$ for $1\le \tau \le \left|V\right|-2$ such that $$A_{ik_1}\prod_{t=1}^{\tau-1} A_{k_{t}k_{t+1}}A_{k_{\tau}j}=1.$$ If there is such a path, we define $\mathcal{E}\left(v_{i}, v_{j}\right)=1$; otherwise, $\mathcal{E}\left(v_{i}, v_{j}\right)=0$. This function is useful for na\"{\i}ve PCE, which uses the network inputs that influence $y$, denoted as $\boldsymbol{\xi} := \boldsymbol{x}_{S(G)\cap V_y}$ for $$V_y:=\left\{v_{i} \in V: \mathcal{E}\left(v_{i}, v_{y}\right)=1 \right\}.$$  
We assume the variables in $\boldsymbol{\xi}$ are mutually independent hereafter to well-define the Sobol indices of $y$ with respect to $\boldsymbol{\xi}$.

For network PCE, which relates each node to its direct predecessors, we define 
\begin{equation}
    \begin{aligned}
    \mathscr{P}\!\left(\boldsymbol{x}_{\boldsymbol{v}}\right) := \boldsymbol{x}_{ \left\{v_{i} \in V: v_{j}\in \boldsymbol{v}, A_{ij}=1\right \} \cup \left(\boldsymbol{v} \cap S(G) \right)} , %\left\{v_{i} \in V: v_{i}\in \boldsymbol{v}, v_{i} \in S(G) \right\}}.
    \end{aligned}\nonumber
\end{equation}
where $\left\{v_{i} \in V: v_{j}\in \boldsymbol{v}, A_{ij}=1\right \}$ represents the direct predecessors, if any, of the nodes in $\boldsymbol{v}$. To well-define $\mathscr{P}\!\left(\boldsymbol{x}_{\boldsymbol{v}}\right)$, $\boldsymbol{v} \cap S(G)$ represents the source nodes in $\boldsymbol{v}$, which do not have any direct predecessors. For $l\in \mathbb{N}$, $\mathscr{P}^{l}(\boldsymbol{x}_{\boldsymbol{v}})$ denotes applying the operator $\mathscr{P}(\cdot)$ on $\boldsymbol{x}_{\boldsymbol{v}}$ $l$ times. 

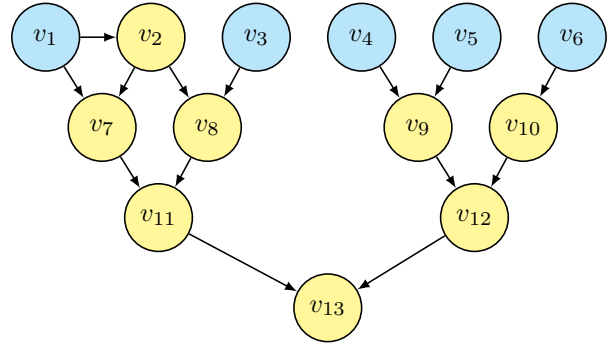
\begin{figure}[H]
{\centering
\begin {tikzpicture}[-latex ,auto ,node distance =1.2cm and 1.4cm ,on grid ,
semithick , state/.style ={ circle ,
draw, fill = cyan!25, text=black , minimum width =0.9 cm}, output/.style = {circle ,  draw, fill = yellow!50 , text=black , minimum width =0.9 cm}]
actor role/.style = {rectangle, draw=black!80, ultra thick,
    minimum size = 5mm, every actor role},
\node[state] (X1) [xshift = 0.1cm]{$v_{1}$ };
\node[output] (X2) [right =of X1] {$v_{2} $};
\node[state] (X3) [right =of X2] {$v_{3} $};
\node[state] (X4) [right =of X3] {$v_{4} $};
\node[state] (X5) [right =of X4] {$v_{5} $};
\node[state] (X6) [right =of X5] {$v_{6} $};
\node[output] (X7) [below =of X1, xshift = 0.75cm] {$v_{7}$};
\node[output] (X8) [right =of X7] {$v_{8}$};
\node[output] (X9) [below =of X4, xshift = 0.75cm] {$v_{9}$};
\node[output] (X10) [right =of X9] {$v_{10}$};
\node[output] (X11) [below =of X7, xshift = 0.75cm] {$v_{11}$};
\node[output] (X12) [below =of X9, xshift = 0.75cm] {$v_{12}$};
\node[output] (Y) [below =of X11, xshift = 2.25cm] {$v_{13}$};
\path (X1) edge [] node[below = 0.15 cm] {}(X2);
\path (X1) edge [] node[below = 0.15 cm] {}(X7);
\path (X2) edge [] node[below = 0.15 cm] {}(X7);
\path (X2) edge [] node[below = 0.15 cm] {}(X8);
\path (X3) edge [] node[below = 0.15 cm] {}(X8);
\path (X4) edge [] node[below = 0.15 cm] {}(X9);
\path (X5) edge [] node[below = 0.15 cm] {}(X9);
\path (X6) edge [] node[below = 0.15 cm] {}(X10);
\path (X7) edge [] node[below = 0.15 cm] {}(X11);
\path (X8) edge [] node[below = 0.15 cm] {}(X11);
\path (X9) edge [] node[below = 0.15 cm] {}(X12);
\path (X10) edge [] node[below = 0.15 cm] {}(X12);
\path (X11) edge [] node[below = 0.15 cm] {}(Y);
\path (X12) edge [] node[below = 0.15 cm] {}(Y);
\end{tikzpicture}
\centering\caption{\small{This example network contains all the four 3-node motifs of DAG. The network inputs represented by the blue-shaded nodes, $x_{v_{1}}, x_{v_{3}}, x_{v_{4}}, x_{v_{5}}$, and $x_{v_{6}}$, are mutually 
independent. The variable represented by node $v_{13}$, $x_{v_{13}}$, is the network output $y$. The arrows represent dependent relationships; e.g., $y=x_{v_{13}}$ directly depends on $x_{v_{11}}$ and $x_{v_{12}}$, while indirectly depending on all the network inputs.}}
\label{fig:DAG}
}
\end{figure}

Network PCE can be applied to any DAG, which contains any of the four possible 3-node motifs \cite{carstens2013motifs}. Fig.~\ref{fig:DAG} shows an example DAG, which contains all the four motifs (e.g., $\{v_1, v_2, v_7\}, \{v_2, v_7, v_8\}, \{v_4, v_5, v_9\}, \{v_6, v_{10}, v_{12}\}$). Note the network inputs $\boldsymbol{\xi}=\left(x_{v_{1}}, x_{v_{3}}, x_{v_{4}}, x_{v_{5}}, x_{v_{6}}\right)$ are mutually independent. %$Y$ is the network output. 
The node $v_{13}$ is the sink node corresponding to the network output $y$.
%The goal is to obtain the Sobol indices of the network output $x_{v_{13}}$ with respect to the network inputs.  
This example DAG will be used in the following subsections to illustrate na\"{\i}ve PCE and network PCE. % to obtain these indices.

\subsection{Na\"{\i}ve PCE}
Na\"{\i}ve PCE is the standard PCE in eq.~\eqref{eq:exp} that approximates $y$ directly as a function of the network inputs that influence $y$, $\boldsymbol{\xi} = \boldsymbol{x}_{S(G)\cap V_y}$, as follows:  
\begin{equation}
    \label{eq:naive_1}
    \hat{y} = \sum_{i=0}^{P}\theta_{i}\psi_{i}(\boldsymbol{\xi}).
\end{equation}
%Here, $P$ is the number of orthonormal polynomial terms that depends on $\dim(\boldsymbol{\xi})$ which is the number of the variables in $\boldsymbol{\xi}$ and the highest polynomial order $p$. 
This PCE directly yields the estimated Sobol indices of $y$ with respect to $\boldsymbol{\xi}$ based on eqs.~\eqref{eq:sobeqs} and \eqref{eq:sobeqs2}. The na\"{\i}ve PCE algorithm is summarized in Algorithm \ref{algorithm_naive}.

%  , the number of the orthonormal polynomials increases exponentially as the number of network inputs increases. Therefore, in order to solve eq.~\eqref{eq:over_lin}, the number of minimally required observations increases exponentially as the number of network inputs increases. 
%  Therefore, the na\"{\i}ve PCE algorithm may only fit for small-scale DAGs, where the number of the inputs in the PCE is not large.  

\begin{algorithm}[!ht]
\caption{Na\"{\i}ve PCE Algorithm}
%\label{}
\begin{algorithmic}[1]\label{algorithm_naive} 
\REQUIRE $G=DAG(V, E)$; at least $P+1$ observations of the network output $y$ and inputs $\boldsymbol{\xi} = \boldsymbol{x}_{S(G)\cap V_y}$. \\%$v_{y} \in V$: the sink node that represents the network output $y$. \\%, highest polynomial order $p$\\
\ENSURE Sobol indices of $y$ with respect to each input in $\boldsymbol{\xi}$. 
%\STATE{Find the network inputs that influence $y$, $\boldsymbol{\xi} = \boldsymbol{x}_{S(G)\cap V_y}$.}
\STATE{Construct univariate orthonormal polynomials for each input in $\boldsymbol{\xi}$.}
\STATE{Construct multivariate orthonormal polynomials for $\boldsymbol{\xi}$ as the tensor products of the univariate orthonormal polynomials using eq.~\eqref{eq:multi}.}
\STATE{Estimate $\boldsymbol{\theta}$ in eq.~\eqref{eq:naive_1} by solving an equivalent problem to eq.~\eqref{eq:over_lin}.}
\STATE{Estimate the Sobol indices based on the estimated $\boldsymbol{\theta}$ using eqs.~\eqref{eq:sobeqs} and \eqref{eq:sobeqs2}}.
\end{algorithmic}
\end{algorithm}

As discussed in Sec.~\ref{subsec:pce}, the number of orthonormal polynomials, $P$, increases exponentially in $\dim\left(\boldsymbol{\xi}\right)$. Thus, this na\"{\i}ve approach of taking the network as a black-box requires an exponentially increasing number of observations to solve an equivalent problem to eq.~\eqref{eq:over_lin} as $\dim\left(\boldsymbol{\xi}\right)$ increases. In other words, a sensitivity analysis with respect to a large number of network inputs requires a large number of observations for na\"{\i}ve PCE. This issue is mitigated by network PCE that explicitly considers the network structure. 
%\noindent Depending on the knowledge of the random inputs, the univariate orthonormal polynomials can be constructed using different PCEs for step 2 in Algorithm 1.

\subsection{Network PCE}\label{subsec:nPCE}
{\color{black}We propose} network PCE to leverage the known network structure of the process of interest to improve the efficiency and accuracy of sensitivity analysis. Intuitively speaking, this model recursively relates a network node to its direct predecessors to trace the output variance back to the network inputs. How uncertainties propagate through the network is effectively captured by the PCE coefficients in the model. The coefficients directly yield estimated Sobol indices of the output with respect to each input in the network. %which recursively building PCE of $y$ with respect to the network inputs by leveraging the network structure.  

The recursive modeling process for network PCE starts from the sink node $v_y$ (e.g., $v_{13}$ in Fig.~\ref{fig:DAG}) and traces it back to the source nodes (e.g., $v_{1}, v_{3}, v_{4}, v_{5}, v_{6}$ in Fig.~\ref{fig:DAG}). The network output $y$ is first modeled as the PCE in $\mathscr{P}(y)$ (e.g., $(x_{v_{11}}, x_{v_{12}})$ in Fig.~\ref{fig:DAG}), which includes the variables corresponding to the direct predecessors of $v_y$. Then, their direct predecessors are recursively found until $y$ is modeled as the PCE in $\boldsymbol{\xi}$, or equivalently, $\mathscr{P}^{L}(y)$, where $L:=\inf\{l\in\mathbb{N}: \mathscr{P}^{l}(y) = \mathscr{P}^{l+1}(y) \}$ denotes the total number of iterations.   

In the $l^{th}$ iteration of this recursive process for $l=1,\ldots,L$, we need to find mutually independent vectors to use in a PCE (recall eq.~\eqref{eq:multi}). Thus, we define the \textit{mutually independent decomposition} of $\mathscr{P}^{l}(y)$ as follows: 
\begin{equation}
    \label{eq:decom}
    %\mathscr{M}\!\left(\mathscr{P}(y)\right):=
    \boldsymbol{x}_{\boldsymbol{v}^{(l)}} := \left(\boldsymbol{x}_{\boldsymbol{v}_{1}^{(l)}}, \boldsymbol{x}_{\boldsymbol{v}_{2}^{(l)}}, \ldots, \boldsymbol{x}_{\boldsymbol{v}_{n^{(l)}}^{(l)}}\right),  
\end{equation}
which has an arbitrary order of the elements and satisfies the following two conditions:  
\begin{enumerate}
    \item $\boldsymbol{x}_{\boldsymbol{v}_{i}^{(l)}} \!\perp\!\!\!\perp \boldsymbol{x}_{\boldsymbol{v}_{j}^{(l)}}, \forall  i\neq j$;
    \item $x_{v_j} \not\!\perp\!\!\!\perp x_{v_k}, \forall v_j, v_k \in \boldsymbol{v}_{i}^{(l)}, \forall i$.
\end{enumerate}
Note $n^{(l)}:= \dim\left(\boldsymbol{x}_{\boldsymbol{v}^{(l)}} \right)$ is the number of elements of the tuple in eq.~\eqref{eq:decom}.
For example, in Fig.~\ref{fig:DAG}, the mutually independent decompositions of $\mathscr{P}(y)$ and  $\mathscr{P}^{2}(y)$ are $\left(x_{v_{11}}, x_{v_{12}}\right)$ and $\left( \left(x_{v_{7}}, x_{v_{8}}\right), x_{v_{9}}, x_{v_{10}}\right)$, respectively. Hence, $n^{(1)}=2$ and $n^{(2)}=3$, not 4.

In the $l^{th}$ iteration, network PCE approximates $y$ using the PCE as follows:  
\begin{equation}
    \label{eq:network_k}
    \hat{y}^{(l)} := \sum_{i=0}^{P^{(l)}}\theta_{i}^{(l)}\psi^{(l)}_{i}\!\left(\mathscr{P}^{l}\!\left(y\right)\right), 
\end{equation}
%where $\hat{y}^{(l)}$ is the PCE approximation of $y$ after $l$ iterations. And $P^{(l)}$ is the number of orthonormal polynomials. 
where each orthonormal polynomial $\psi^{(l)}_{i}\!\left(\mathscr{P}^{l}\!\left(y\right)\right)$ can be obtained using the mutually independent decomposition of $\mathscr{P}^{l}(y)$ in eq.~\eqref{eq:decom} as follows:
\begin{equation}
    \label{eq:vec_tensor_l}
    \psi^{(l)}_{i}\left(\mathscr{P}^{l}(y)\right)= \psi^{(l)}_{\boldsymbol{\alpha}_{i}^{(l)}}\left(\mathscr{P}^{l}(y)\right) := \prod_{j=1}^{n^{(l)}}\psi_{\boldsymbol{\alpha}_{ij}^{(l)}}^{(l)}\left(\boldsymbol{x}_{\boldsymbol{v}_{j}^{(l)}}\right).
\end{equation}
Here $\boldsymbol{\alpha}_{i}^{(l)}=\left(\boldsymbol{\alpha}_{i1}^{(l)}, \ldots, \boldsymbol{\alpha}_{in^{(l)}}^{(l)}\right)$ is the $i$-th arbitrary tuple satisfying $\sum_{j=1}^{n^{(l)}}\left|\boldsymbol{\alpha}_{ij}^{(l)}\right| \leq p^{(l)}$ for the pre-specified highest polynomial order $p^{(l)}$. $\boldsymbol{\alpha}_{ij}^{(l)}$ is the vector composed of the polynomial orders with respect to the variables in $\boldsymbol{x}_{\boldsymbol{v}_{j}^{(l)}}$ for $j=1,2,\ldots,n^{(l)}$. $\psi_{\boldsymbol{\alpha}_{ij}^{(l)}}^{(l)}\left(\boldsymbol{x_{\boldsymbol{v}_{j}^{(l)}}}\right)$ is an $\left|\boldsymbol{\alpha}_{ij}^{(l)}\right|$-th order orthonormal polynomial in $\boldsymbol{x_{\boldsymbol{v}_{j}^{(l)}}}$ obtained using the modified Gram-Schmidt algorithm \cite{liu2018data}. %based on arbitrary ordering of the polynomials in $\boldsymbol{x_{\boldsymbol{v}_{j}^{(l)}}}$ \cite{liu2018data}.  

In the first iteration ($l=1$), the PCE coefficients, $\theta_{i}^{(l)}, i=0,1,\ldots, P^{(l)}$, in eq.~\eqref{eq:network_k} are estimated by solving an equivalent problem to eq.~\eqref{eq:over_lin} where only the notations are different. For the subsequent iterations ($l\ge 2$), the coefficients are calculated in a different way using the previous iteration's coefficients, as detailed next. 

To model $y$ as a function of $\mathscr{P}^{l+1}(y)$, or equivalently, 
$$
\left(\mathscr{P}\!\left(\boldsymbol{x}_{\boldsymbol{v}_{1}^{(l)}}\right), \mathscr{P}\!\left(\boldsymbol{x}_{\boldsymbol{v}_{2}^{(l)}}\right), \ldots, \mathscr{P}\!\left(\boldsymbol{x}_{\boldsymbol{v}_{n^{(l)}}^{(l)}}\right)\right),
$$
network PCE substitutes $\psi_{\boldsymbol{\alpha}_{ij}^{(l)}}^{(l)}\left(\boldsymbol{x_{\boldsymbol{v}_{j}^{(l)}}}\right)$ in eq.~\eqref{eq:vec_tensor_l} %using the variables corresponding to the direct predecessors of $\boldsymbol{v}_{j}^{(l)}$ 
with 
\begin{equation}
\begin{aligned}
    \label{eq:iter2}
    \hat{\psi}_{\boldsymbol{\alpha}_{ij}^{(l)}}^{(l)}\left(\boldsymbol{x}_{\boldsymbol{v}_{j}^{(l)}}\right) := \sum_{k=0}^{P_{ij}^{(l)}}\theta_{ijk}^{(l)}\psi_{ijk}^{(l)}\left(\mathscr{P}\!\left(\boldsymbol{x}_{\boldsymbol{v}_{j}^{(l)}}\right)\right)
\end{aligned}    
\end{equation}
to obtain the approximation of eq.~\eqref{eq:vec_tensor_l} as follows:
\begin{equation}
\begin{aligned}
\label{eq:iter1}
&\hat{\psi}^{(l)}_{i}\left(\mathscr{P}^{l}(y)\right) := \prod_{j=1}^{n^{(l)}}\hat{\psi}_{\boldsymbol{\alpha}_{ij}^{(l)}}^{(l)}\left(\boldsymbol{x}_{\boldsymbol{v}_{j}^{(l)}}\right).
\end{aligned}
\end{equation}
Without loss of generality, we let the highest polynomial order $p_{ij}^{(l)}$ of orthonormal polynomials in eq.~\eqref{eq:iter2} to be the constant $p^{(l+1)}$. 
Substituting $\hat{\psi}_{i}^{(l)}\left(\mathscr{P}^{l}(y)\right)$ in eq.~\eqref{eq:iter1} for $\psi^{(l)}_{i}\!\left(\mathscr{P}^{l}\!\left(y\right)\right)$ in eq.~\eqref{eq:network_k} yields the approximation of $y$  for the $(l+1)^{th}$ iteration as follows:
\begin{align}
    \hat{y}^{(l+1)} &= \sum_{i=0}^{P^{(l)}}\theta_{i}^{(l)}\hat{\psi}^{(l)}_{i}\left(\mathscr{P}^{l}(y)\right) \label{eq:inductive_k+1} \\ 
    &:= \sum_{i=0}^{P^{(l+1)}}\theta_{i}^{(l+1)}\psi_{i}^{(l+1)}\left(\mathscr{P}^{l+1}(y)\right), \label{eq:nk_l}
\end{align}
where the PCE coefficient $\theta_{i}^{(l+1)}$ in eq.~\eqref{eq:nk_l} is calculated after estimating the coefficients in eq.~\eqref{eq:iter2} (by solving an equivalent problem to eq.~\eqref{eq:over_lin}) and rearranging the terms in eq.~\eqref{eq:inductive_k+1}, which involve the previous iteration's coefficients. This recursive approach of calculating the PCE coefficient helps reduce the minimally required number of observations compared to na\"{\i}ve PCE, as explained later. 

The above iteration ends when $y$ is approximated as the PCE in $\mathscr{P}^{L}(y)$, or equivalently, $\boldsymbol{\xi}$ as follows: 
\begin{equation}
    \label{eq:nk_f}
    \hat{y}^{(L)} := \sum_{i=0}^{P^{(L)}}\theta_{i}^{(L)}\psi_{i}(\boldsymbol{\xi}),
\end{equation}
where $P^{(L)}$ depends on the highest polynomial orders, $p^{(1)}, \ldots, p^{(L)}$. This recursive approximation process is valid as shown in Theorem~\ref{thm:npCE} below, which proves the convergence of $\hat{y}^{(L)}$ in eq.~\eqref{eq:nk_f} to $y$ in probability. The PCE coefficients, $\theta_{i}^{(L)}, i=1,\ldots,P^{(L)}$, in eq.~\eqref{eq:nk_f} directly yield the estimated Sobol indices of $y$ with respect to each input in $\boldsymbol{\xi}$, as described in Sec.~\ref{subsec:PCE_Sobol}. The network PCE algorithm is summarized in Algorithm~\ref{algorithm_nPCE}.

\begin{algorithm}[ht]
\caption{Network PCE Algorithm}
%\label{}
\begin{algorithmic}[1]\label{algorithm_nPCE} 
\REQUIRE $G=DAG(V, E)$; at least $$\max\left(P^{(1)}, \underset{\substack{1\le l \le L-1 \\ 0\le i \le P^{(l)} \\ 1\le j \le n^{(l)}}  }{\max}P_{ij}^{(l)} \right)+1$$ observations of the network output $y$ and all the variables in $\boldsymbol{x}_{V_y}$; Iteration counter $l=1$. \\%,  $v_{y} \in V$: the sink node that represents the network output $y$.\\
\ENSURE Sobol indices of $y$ with respect to each input in $\boldsymbol{\xi}= \boldsymbol{x}_{S(G)\cap V_y}$. 
%\STATE{Find the mutually independent decomposition of $\mathscr{P}^{l}(y)$ in eq.~\eqref{eq:decom}.}
\STATE{Construct  $\psi_{\boldsymbol{\alpha}_{ij}^{(l)}}^{(l)}\left(\boldsymbol{x}_{\boldsymbol{v}_{j}^{(l)}}\right)$ in eq.~\eqref{eq:vec_tensor_l} using the modified Gram-Schmidt algorithm for $i=0,1,\ldots,P^{(l)}$ and $j= 1,2,\ldots, n^{(l)}$.}% based on an arbitrary order of the polynomials in $\boldsymbol{x}_{\boldsymbol{v}_{j}^{(l)}}$.}
\STATE{Construct $\psi^{(l)}_{i}\!\left(\mathscr{P}^{l}\!\left(y\right)\right)$ for $i=0,1,\ldots,P^{(l)}$   in eq.~\eqref{eq:network_k} as the tensor product of $\psi_{\boldsymbol{\alpha}_{ij}^{(l)}}^{(l)}\left(\boldsymbol{x}_{\boldsymbol{v}^{(l)}_{j}}\right)$ for $j= 1,2,\ldots, n^{(l)}$ in eq.~\eqref{eq:vec_tensor_l}.}
\STATE{If $l=1$, estimate the PCE coefficients, $\theta_{i}^{(l)}, i=0,1,\ldots, P^{(l)}$, in eq.~\eqref{eq:network_k} by solving an equivalent problem to eq.~\eqref{eq:over_lin}. If $L=1$, skip to Step 6.}
\STATE{Estimate the PCE coefficients in eq.~\eqref{eq:iter2} and substitute eq.~\eqref{eq:iter1} into eq.~\eqref{eq:inductive_k+1}.}
\STATE{Rearrange the terms in eq.~\eqref{eq:inductive_k+1} to yield the expression in eq.~\eqref{eq:nk_l}. Increment $l$ by 1 and if $l<L$, go to Step 1.}
\STATE{Estimate the Sobol indices using eqs.~\eqref{eq:sobeqs} and \eqref{eq:sobeqs2} based on the PCE coefficients $\theta_{i}^{(L)}, i=0,1,\ldots, P^{(L)}$, in eq.~\eqref{eq:nk_f}.}
\end{algorithmic}
\end{algorithm}

\begin{figure*}[!ht]
{\centering
\includegraphics[width=0.75\textwidth, height= 5.5cm]{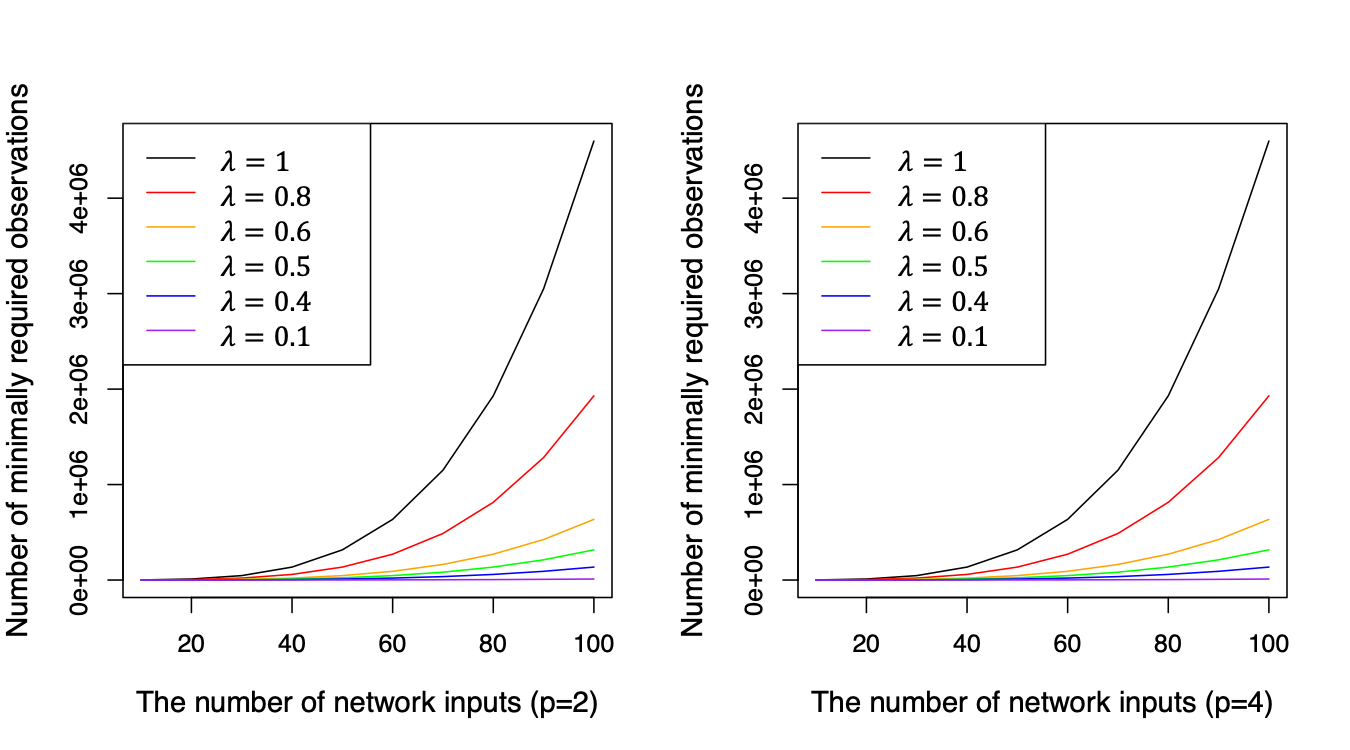} \\
\centering\caption{\small{Network PCE ($\lambda < 1$) requires fewer observations to model the network output than na\"{\i}ve PCE ($\lambda = 1$) for a larger $p$ (the highest polynomial order used in both PCEs) and a smaller $\lambda$. $\lambda$, defined in eq.~\eqref{eq:lambda}, represents the ratio of the largest number of variables used in any PCE for network PCE to the number of network inputs used in na\"{\i}ve PCE.}}
\label{fig:eff}
}
\end{figure*}

An advantage of network PCE over na\"{\i}ve PCE lies in its potential %counter example: single network input & many intermediate nodes
to use much fewer observations for solving equivalent problems of eq.~\eqref{eq:over_lin}. The minimally required number of observations for either model is equal to the largest number of orthonormal polynomials of any PCE used in the model. Thanks to the recursive decomposition procedure of network PCE, it tends to use a much smaller PCE than na\"{\i}ve PCE especially when the number of the inputs that influence $y$ is large (i.e., $\dim\!\left(\boldsymbol{\xi}\right) \gg 1$). To illustrate this point, %consider a network where all of its inputs influence the output of interest (i.e., $S(G)\subseteq V_y$), and 
assume all PCEs in network PCE use the same highest polynomial order $p^{(l)} = p$, $l=1,2,\ldots, L$, as na\"{\i}ve PCE. Then, the minimally required number of observations for na\"{\i}ve PCE increases exponentially in $\dim\!\left(\boldsymbol{\xi}\right)$, as explained below eq.~\eqref{eq:exp}. That for network PCE (see Input of Algorithm~\ref{algorithm_nPCE}) increases exponentially in the largest number of variables used for any PCE in Steps 3 and 4 of Algorithm~\ref{algorithm_nPCE}, 
\begin{equation}
    \label{eq:num_var_nPCE}
\max\left(\left| D_y \right|, \underset{1\le l \le L-1}{\mathrm{max}} \underset{1\le j\le n^{(l)}}{\mathrm{max}} \left|D_j^{(l)}\right|\right).
\end{equation} 
Here,
\begin{equation}
    \label{eq:direct_pred_y}
D_y := \{v_i \in V: \langle v_{i}, v_{y} \rangle \in E   \}
\end{equation}
denotes the set of all direct predecessors of the output node $v_y$ and
\begin{equation}
    \label{eq:direct_pred_jl}
    D_j^{(l)} :=  \{v_i \in V: v_k \in \boldsymbol{v}_{j}^{(l)}, \langle v_{i}, v_{k} \rangle \in E   \}  
\end{equation}
denotes the set of all direct predecessors of the nodes in $\boldsymbol{v}_{j}^{(l)}$ in eq.~\eqref{eq:iter2}. 
Fig.~\ref{fig:eff} visualizes the minimally required number of observations with respect to $p$ and the ratio
\begin{equation}
    \label{eq:lambda}
\lambda := \frac{\max\left(\left| D_y \right|, \underset{1\le l \le L-1}{\mathrm{max}} \underset{1\le j\le n^{(l)}}{\mathrm{max}} \left|D_j^{(l)}\right|\right)}{\dim\!\left(\boldsymbol{\xi}\right)}.
\end{equation}
The saving of network PCE (with $\lambda < 1$) over na\"{\i}ve PCE ($\lambda =1$) increases  as $p$ increases or $\lambda$ decreases.        

\subsection{Sparse network PCE (SN-PCE)}
{\color{black}While the proposed network PCE  already provides substantial parsimony over na\"{\i}ve PCE, it may still need to use} many orthonormal polynomials to capture all the main effects and the major interaction effects of the network variables on the output variance. However, in practice, many of such effects are insignificant. Therefore, we propose SN-PCE, which imposes $l_{1}$-sparsity on the PCE coefficients to capture only significant pathways of uncertainty propagation in the network. Specifically, SN-PCE solves an equivalent problem of eq.~\eqref{eq:lasso} (with the difference lying only in the notations) instead of eq.~\eqref{eq:over_lin} in Steps 3 and 4 of Algorithm~\ref{algorithm_nPCE}. The resulting parsimonious model can use even fewer observations than network PCE to estimate the PCE coefficients and, in turn, the Sobol indices. Also, because only significant effects appear in the model (with non-zero PCE coefficients), the model is easier to interpret than the other models.

{\color{black}
\subsection{Theoretical result}
}
Since na\"{\i}ve PCE is a direct application of standard PCE in eq.~\eqref{eq:exp}, the corresponding theoretical result (Theorem~\ref{thm:naive}) directly follows from \cite{rahman2018polynomial} by appropriately invoking Assumption \ref{ass_3} below. The result is still presented here mainly for comparison with the theoretical result on network PCE (Theorem~\ref{thm:npCE}).

\begin{assumption}[adapted from \cite{rahman2018polynomial}]
\label{ass_3}
 The random vector $\boldsymbol{x} \coloneqq (x_{1}, \ldots, x_{n})^{T}: (\Omega, \mathcal{F}) \rightarrow \left(\mathbb{A}^{n}, \mathcal{B}^{n}\right)$
\begin{enumerate}

    \item has a continuous joint probability density function $\mu(\boldsymbol{x})$ with a bounded or unbounded support $\mathbb{A}^{n} \subseteq \mathbb{R}^{n}$; 
    \item possesses absolute finite moments of all orders, that is, $\forall \mathbf{j}\coloneqq (j_{1}, j_{2}, \ldots, j_{n}) \in \mathbb{N}_{0}^{n}$, 
\[
\mathbb{E}\left(\left|\boldsymbol{x}^{\mathbf{j}}\right|\right) := \int_{\Omega}\left|\boldsymbol{x}^{\mathbf{j}}(w)\right|d\mathcal{P}(w) <\infty, 
\]
where $\boldsymbol{x}^{\mathbf{j}} := x_{1}^{j_{1}}\cdots x_{n}^{j_{n}}$; % and $\mathbb{E}$ is the expectation operator with respect to the probability measure $\mathcal{P}$ or $\mu(\boldsymbol{x})d\boldsymbol{x}$; 
and
\item has a joint probability density function $\mu(\boldsymbol{x})$, which 
\begin{enumerate}
    \item has a compact support, that is, there exists a compact subset $\mathbb{A}^{n} \subset \mathbb{R}^{n}$ such that $\mathcal{P}(\boldsymbol{x} \in \mathbb{A}^{n}) = 1$, or
    \item is exponentially integrable, that is, there exists a real number $\alpha >0$ such that 
    \[
    \int_{\mathbb{A}^{n}}\exp\left(\alpha||\boldsymbol{x}||\right)\mu(\boldsymbol{x})\,d\boldsymbol{x} < \infty,
    \]
      where $||\cdot||: \mathbb{A}^{n} \rightarrow \mathbb{R}_{0}^{+}$ is an abitrary norm. 
\end{enumerate}
\end{enumerate}
\end{assumption}

\begin{theorem}\label{thm:naive}
Suppose that %Assumptions \ref{ass_1} and \ref{ass_2} hold for $G=DAG(V,E)$ and that 
for the network output $y\!\left(\boldsymbol{\xi}\right) \in \mathcal{L}^{2}\!\left(\Omega, \mathcal{F}, \mathcal{P}\right)$, $\boldsymbol{\xi} = \boldsymbol{x}_{S(G)\cap V_y}$ fulfills Assumption \ref{ass_3}. Then, $\hat{y}$ in eq.~\eqref{eq:naive_1} converges to $y$ in probability as $P \to \infty$.

\end{theorem}

\begin{proof}
This is a direct result of Theorem 11 in \cite{rahman2018polynomial}.
\end{proof}

Theorem~\ref{thm:naive} imposes regularity conditions on the network inputs $\boldsymbol{\xi}$ and output $y$, but not on the other network variables. This black-box approach of na\"{\i}ve PCE still provides a convergent prediction of $y$ as $P\to\infty$. In practice, increasing $P$ requires increasing the number of observations. Thus, the finite-sample inefficiency of na\"{\i}ve PCE compared to network PCE (described in Sec.~\ref{subsec:nPCE}) is critical when the sample size is limited.% due to resource constraints.

Theoretical results on network PCE impose regularity conditions on all the network variables as the knowledge of network structure provides network PCE's advantage over na\"{\i}ve PCE. Lemma~\ref{lemma_2} validates the recursive modeling of a network node using its direct predecessors (described in Steps 1--5 in Algorithm~\ref{algorithm_nPCE}). Building on Lemma~\ref{lemma_2}, Theorem~\ref{thm:npCE} proves the convergence of network PCE output in eq.~\eqref{eq:nk_f}.

\begin{lemma}
\label{lemma_2}
Suppose that $\boldsymbol{x}_{\boldsymbol{v}_{j}^{(l)}}$ is a function of $\mathscr{P}\!\left(\boldsymbol{x}_{\boldsymbol{v}_{j}^{(l)}}\right)$ for $l= 1, \ldots,L-1$ and $j= 1,\ldots, n^{(l)}$,  %Assumptions \ref{ass_1} and \ref{ass_2} hold for $G= DAG(V,E)$ 
and that for the network output $y$, $\boldsymbol{x}_{V_y}$  fulfills Assumption \ref{ass_3}.  
Then, 
\begin{enumerate}
\item $\hat{\psi}_{\boldsymbol{\alpha}_{ij}^{(l)}}^{(l)}\left(\boldsymbol{x}_{\boldsymbol{v}_{j}^{(l)}}\right)$ in eq.~\eqref{eq:iter2} converges to $\psi_{\boldsymbol{\alpha_{ij}}^{(l)}}^{(l)}\left(\boldsymbol{x}_{\boldsymbol{v}_{j}^{(l)}}\right)$ in eq.~\eqref{eq:vec_tensor_l} in probability as $P_{ij}^{(l)} \to \infty$; and 
\item $\hat{\psi}_{i}^{(l)}\left(\mathscr{P}^{l}\left(y\right)\right) = \prod_{j=1}^{n^{(l)}}\hat{\psi}_{\boldsymbol{\alpha}_{ij}^{(l)}}^{(l)}\left(\boldsymbol{x}_{\boldsymbol{v}_{j}^{(l)}}\right)$ in eq.~\eqref{eq:iter1} converges to $\psi^{(l)}_{i}\left(\mathscr{P}^{l}(y)\right)$ in eq.~\eqref{eq:vec_tensor_l} in probability as $P_{ij}^{(l)} \to \infty$
\end{enumerate}
for all $l= 1, \ldots,L-1$ and all $\boldsymbol{\alpha}_{i}^{(l)}=\left(\boldsymbol{\alpha}_{i1}^{(l)}, \ldots, \boldsymbol{\alpha}_{in^{(l)}}^{(l)}\right)$ satisfying $\sum_{j=1}^{n^{(l)}}\left|\boldsymbol{\alpha}_{ij}^{(l)}\right| \leq p^{(l)}$. 
\begin{proof}
%By Assumption~\ref{ass_3},
Under the stated assumptions, $\psi_{\boldsymbol{\alpha_{ij}}^{(l)}}^{(l)}\left(\boldsymbol{x}_{\boldsymbol{v}_{j}^{(l)}}\right)$ in eq.~\eqref{eq:vec_tensor_l} is a square-integrable function of $\mathscr{P}\!\left(\boldsymbol{x}_{\boldsymbol{v}_{j}^{(l)}}\right)$. %YC: because the exponential integrability in Assumption~\ref{ass_3} ensures that any squared polynomials are integrable.
The first statement follows from Theorem 11 in \cite{rahman2018polynomial}. The second statement follows from the first statement by the continuous mapping theorem.  
\end{proof}

\end{lemma}

\begin{theorem}\label{thm:npCE} Suppose that the assumptions in Lemma~\ref{lemma_2} hold for the network output $y\!\left( \boldsymbol{x}_{\boldsymbol{v}^{(1)}} \right)\in \mathcal{L}^{2}\!\left(\Omega, \mathcal{F}, \mathcal{P}\right)$.
Then, there exists an increasing function $\gamma^{(l)}\!: \mathbb{N}\mapsto \mathbb{N}$ such that if $p^{(l+1)} = \gamma^{(l)}\!\left(P^{(l)}\right)$ for $l=1,\ldots,L-1$,
$\hat{y}^{(L)}$ in eq.~\eqref{eq:nk_f} converges to $y$ in probability as $p^{(1)}\to \infty$, or equivalently, $P^{(1)} \to \infty$.
\end{theorem}
\begin{proof}
The proof is provided in the article's online supplementary document. 
\end{proof}

The increasing function $\gamma^{(l)}$ in Theorem~\ref{thm:npCE} prescribes how fast the highest polynomial order $p^{(l+1)} = \gamma^{(l)}\!\left(P^{(l)}\right)$ should grow in relation to the number of orthonormal polynomials, $P^{(l)}$, for $l=1,\ldots,L-1$ so that $\hat{y}^{(L)}$ in eq.~\eqref{eq:nk_f} converges to $y$ in probability. Furthermore, to make the relation between $p^{(l+1)}$ and $p^{(l)}$ more explicit, using eqs.~\eqref{eq:direct_pred_y} and \eqref{eq:direct_pred_jl}, let $d^{(1)} := \left| D_y \right|$ and $d^{(l)} :=\left|\underset{1\le j\le n^{(l)}}{\bigcup} D_j^{(l-1)}\right|$, $l=2,\ldots,L$, denote the number of nodes represented by $\mathscr{P}\!\left(y\right)$ and $\mathscr{P}^{l}\!\left(y\right)$ in eq.~\eqref{eq:network_k}, respectively. Because 
$$P^{(l)}+1 = \binom{d^{(l)}+p^{(l)}}{d^{(l)}},$$% \;\textrm{and}\; P^{(l)}+1 \le {{ d^{(l)}+p^{(l)}} \choose {d^{(l)}}}$$
%$= {{ n^{(l)}+p^{(l)}} \choose {n^{(l)}}}$ 
for $l=1,\ldots,L$,  %eq.~\eqref{eq:nk_l}, 
%The equality holds if $\mathscr{P}^{l}\!\left(y\right) = \mathscr{P}\!\left(\boldsymbol{x}_{\boldsymbol{v}_{j}^{(l-1)}}\right)$ for any $j$? 
%for the network PCE, %as explained below eq. \eqref{eq:exp}, 
%where $n^{(l)}$ is the fixed number of elements in $\boldsymbol{x}_{\boldsymbol{v}^{(l)}}$ for $l=1,\ldots,L$, 
Stirling's approximation %see eq.(3) in http://page.mi.fu-berlin.de/shagnik/notes/binomials.pdf
yields  
$$p^{(l+1)} = \gamma^{(l)}\!\left(O\!\left(\left(p^{(l)}\right)^{d^{(l)}}\right)\right)$$
using the big O notation.

For example, %note: consider an example based on the empirical finding of exponential convergence rate. 
if $\gamma^{(l)}$ is linear (or superlinear), then $p^{(l+1)}$ should grow as fast as (or faster than)  $d^{(l)}$-th power of $p^{(l)}$, or equivalently, 
$\prod_{l'=1}^{l}d^{(l')}$-th power of $p^{(1)}$ for $l=1,\ldots,L-1$. Intuitively speaking, 
%to control the error of approximating $y$ with $\hat{y}^{(L)}$ in eq.~\eqref{eq:nk_f}, the highest polynomial order $p^{(l)}$ should increase exponentially (or faster) as the iteration $l$ increases up to $L$ that represents the network depth. In other words, 
we should control the approximation errors for the upstream nodes in the network more tightly to ensure that the approximation errors for the downstream nodes (especially the output node) are arbitrarily small. We  note that it is beyond the scope of this paper to establish more detailed characteristics of  $\gamma^{(l)}$.

%While we show the existence of $\gamma^{(l)}$ under certain assumptions, 
SN-PCE, which tends to use much smaller $P^{(l)}, l=1,\ldots,L$, than network PCE, mitigates the need for rapidly increasing $p^{(l)}, l=1,\ldots,L$, while maintaining a similar approximation accuracy.

%%%%%%%%%%%%%%%%%%%%%%%%%%%%%%%%%%%%%%%%%%%%%%%%%%%%%%%%%%%%%%%%%%%%%%%%%%%%%%
% Methodology
%%%%%%%%%%%%%%%%%%%%%%%%%%%%%%%%%%%%%%%%%%%%%%%%%%%%%%%%%%%%%%%%%%%%%%%%%%%%%%
\vspace{-0.3cm}
\section{Applications}\label{sec:appli}
For empirical evaluation of the proposed methodology, we conduct sensitivity analysis of two manufacturing processes {\color{black}(welding and injection molding) in \cite{Nannapaneni:2014}.} %The two manufacturing processes, namely, the welding process and the injection molding process, both have multiple variables whose relationships are modeled as a network in \cite{Nannapaneni:2014}. 
We use the same modeling equations and notations therein (hence, some notation conflicts arise although their meanings are clear from the context). %The flooding process is modeled as a network based on the relationships presented in \cite{iooss:2015}.
Note that the proposed methodology leverages the known directional relationships between inputs and the output (expressed as a DAG), not modeling equations, which are often unknown in practice.

{\color{black}We compare four methods, namely, random sampling \cite{owen2013better}, orthogonal array sampling \cite{tissot2015randomized},} na\"{\i}ve PCE, and SN-PCE, for estimating the first-order and total Sobol indices. The estimation is replicated 50 times to calculate the sample mean and standard error of the estimated Sobol indices for each method.  
%The estimation is based on randomly generated observations of the process variables using the models in \cite{Nannapaneni:2014}. The estimation is replicated 50 times to calculate the sample mean and standard error of the estimated Sobol indices for each method.  

%\cite{owen2013better} 
%\cite{tissot2015randomized}

% For empirical evaluation of the proposed methodology, we conduct sensitivity analysis of two manufacturing processes, namely, the welding process and the injection molding process. Each process has multiple variables whose relationships are modeled as a network in \cite{Nannapaneni:2014}. We use the same modeling equations and notations therein (hence, some notation conflicts arise although their meanings are clear from the context). 

% We compare three methods, a state-of-the-art Monte Carlo method \cite{owen2013better}, the na\"{\i}ve PCE, and the SN-PCE, for estimating the first-order and total Sobol indices. The estimation is based on randomly generated observations of the process variables using the models in \cite{Nannapaneni:2014}. The estimation is replicated 200 times to calculate the sample mean and standard error of the estimated Sobol indices for each method.  

As discussed in Sec.~\ref{subsec:Sobol_ind}, the Sobol indices measure the influence of each network input on the variance of the network output. Henceforth, we call some network inputs \textit{influential inputs} if their first-order Sobol indices are $10^{-1}$ or larger; in other words, the influential inputs explain 10\% or more of the output variance without considering their interactions with the other inputs.

%The Monte Carlo method in \cite{owen2013better} is used as a benchmark method because it is known to be more accurate than other Monte Carlo methods for estimating small Sobol indices. 
We use a large sample for {\color{black}random and orthogonal array sampling}, namely, 10,000 observations {\color{black}unless specified otherwise}. {\color{black}We treat the sample mean from 50 replications using the orthogonal array sampling} as the ground truth if the standard error is below 1\% of the sample mean. Even with a large sample, {\color{black}such Monte Carlo sampling} methods tend to have large standard errors for estimating small \textit{total} Sobol indices \cite{myshetskaya2008monte,owen2013better}, as also observed in this study.  

% from the Monte Carlo method with  orthogonal array sampling of observations and treat its sample mean from the 50 replications as the ground truth if the standard error is below 1\% of the sample mean. Even with a large sample, Monte Carlo methods tend to have large standard errors for estimating small \textit{total} Sobol indices \cite{myshetskaya2008monte,owen2013better}, as also observed in this study. 

%The estimated Sobol indices for the influential inputs using the Monte Carlo method with a large sample of observations are considered as the ground truths \cite{yang2011convergence}. 

\subsection{Welding process}
\begin{figure}[!ht]
{\centering
\begin {tikzpicture}[-latex ,auto ,node distance =1.4cm and 1.0cm ,on grid ,
semithick , state/.style ={ circle ,
draw, fill = cyan!25, text=black , minimum width =0.9 cm}, output/.style = {circle ,  draw, fill = yellow!50 , text=black , minimum width =0.9 cm}]
actor role/.style = {rectangle, draw=black!80, ultra thick,
    minimum size = 5mm, every actor role},
\node[state] (e) [xshift =0.1cm]{$e$ };
\node[state] (g) [right =of e] {$g$};
\node[state] (h) [right =of g] {$h$};
\node[state] (l) [right =of h] {$l$};
\node[state] (t) [right =of l] {$t$};
\node[state] (L) [right =of t] {$L$};
\node[output] (V) [below =of h, xshift = 0.65cm] {$V$};
\node[state] (rho) [right =of V] {$\rho$};
\node[state] (Cp) [right =of rho] {$C_{\rho}$};
\node[state] (Ti) [right =of Cp] {$T_{i}$};
\node[state] (Tf) [right =of Ti] {$T_{f}$};
\node[state] (H) [right =of Tf] {$H$};
\node[output] (E) [below =of Cp, xshift = 0.7cm] {$E$};
\path (e) edge [] node[below = 0.15 cm] {}(V);
\path (g) edge [] node[below = 0.15 cm] {}(V);
\path (h) edge [] node[below = 0.15 cm] {}(V);
\path (l) edge [] node[below = 0.15 cm] {}(V);
\path (t) edge [] node[below = 0.15 cm] {}(V);
\path (L) edge [] node[below = 0.15 cm] {}(V);
\path (V) edge [] node[below = 0.15 cm] {}(E);
\path (rho) edge [] node[below = 0.15 cm] {}(E);
\path (Cp) edge [] node[below = 0.15 cm] {}(E);
\path (Ti) edge [] node[below = 0.15 cm] {}(E);
\path (Tf) edge [] node[below = 0.15 cm] {}(E);
\path (H) edge [] node[below = 0.15 cm] {}(E);
\end{tikzpicture}
\centering\caption{\small{This DAG represents the relationships among the variables in the welding process \cite{Nannapaneni:2014}; the weld volume $V$ depends on six welding parameters (weld zone dimensions: $e$, $g$, $h$, $l$, and $t$; weld length $L$), and the total energy $E$ depends on $V$ and additional parameters, $\rho$, $C_{p}$, $T_{i}$,  $T_{f}$, and $H$. }}
\label{fig:weld_process} 
}
\end{figure}
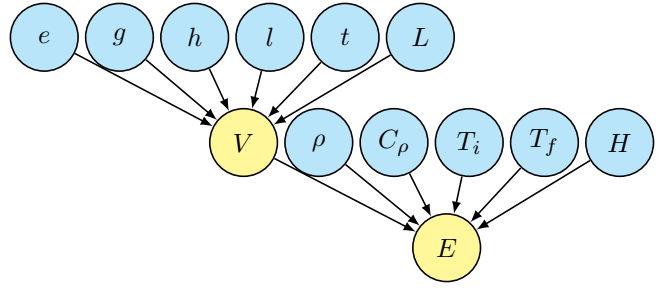
%We consider the welding process in \cite{Nannapaneni:2014} as our first empirical application. 
The welding process has several process variables whose relationships are depicted in Fig.~\ref{fig:weld_process}. The process output of interest is the total minimum theoretical energy required for the welding process, $E$. This energy depends on the weld volume $V$, specific gravity $\rho$, heat capacity $C_{p}$, initial temperature $T_{i}$, final temperature $T_{f}$, and latent heat $H$ as follows:
\begin{equation}
\label{eq:21}
E = \rho V\left( C_{p}(T_{f} - T_{i}) + H \right).\nonumber
\end{equation}
In turn, the weld volume $V$ depends on six welding parameters (weld zone dimensions: $e$, $g$, $h$, $l$, and $t$; weld length $L$) as follows: 
\begin{equation}
\label{eq:20}
V = L\left(0.75lh + gt +0.5(l-9)(t-e)\right).\nonumber
\end{equation}
The process inputs' distributions are presented in Table \ref{tab:sobol-sim}.

\begin{figure*}[!ht]
{\centering
\includegraphics[width=0.4\textwidth]{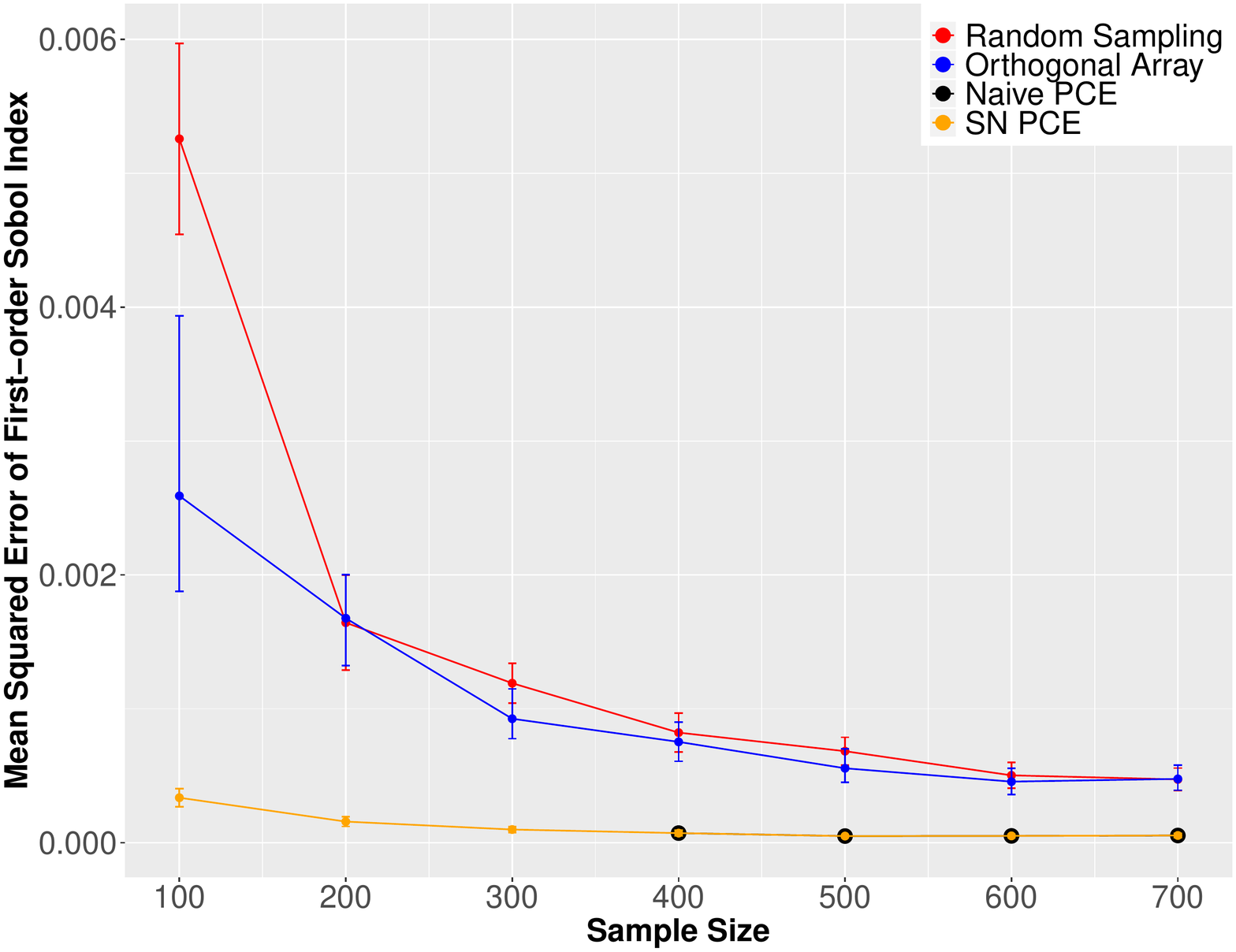}
\includegraphics[width=0.4\textwidth]{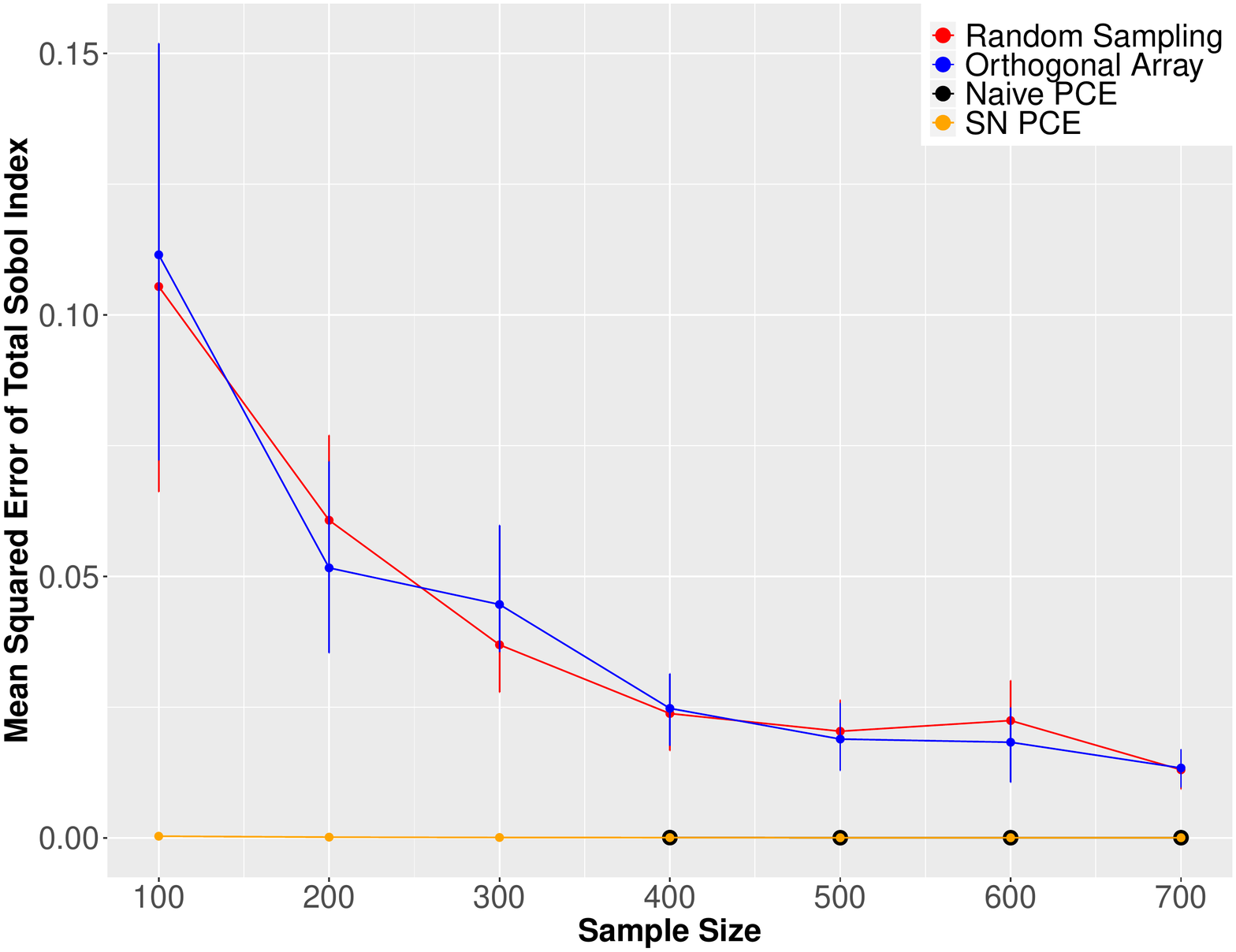} \\
\centering\caption{\small{{\color{black}SN-PCE yields a smaller mean squared error of estimating the first-order Sobol indices (left) and the total Sobol indices (right) than the three other methods for the welding process. Also, SN-PCE requires much fewer observations than na\"{\i}ve PCE to estimate the Sobol indices.}}}
\label{fig:weld_res}
}
\end{figure*}

{\color{black}
We first compare the four methods with respect to the mean squared errors of estimating the Sobol indices when using the same sample size. Fig.~\ref{fig:weld_res} shows that SN-PCE significantly outperforms random and orthogonal array sampling. Also, SN-PCE estimates the Sobol indices using a much smaller sample size than na\"{\i}ve PCE.
%In this experiment, we compared the performance of our proposed SN-PCE method with three different estimators including the Monte Carlo method using random samples, Monte Carlo method using orthogonal array samples, and the na\"{\i}ve PCE. As it is shown in Fig.~\ref{fig:weld_res}, the SN-PCE performances better than the Monte Carlo based estimators with respect to the mean squared errors of estimating Sobol indices for different sample sizes. }
In this example,} we use the highest polynomial order of $3$ for both na\"{\i}ve PCE and SN-PCE; na\"{\i}ve PCE requires at least $364 =\binom{11+3}{3}$ observations for $\dim\!\left(\boldsymbol{\xi}\right)=11$ %\left|S(G)\right|=11
and network PCE (i.e., SN-PCE without sparsity) requires at least $84 = \binom{6+3}{3}$ observations because eq.~\eqref{eq:num_var_nPCE} equals 6. SN-PCE could use even fewer observations because it identifies only $14$ orthonormal polynomials (4\% of those used in na\"{\i}ve PCE) as necessary to approximate the network output of this particular process across all the 50 replications. %{\color{black} As it is shown in Fig.~\ref{fig:weld_res}, compared with the na\"{\i}ve PCE, the SN-PCE can estimate the Sobol indices when only small sample of observations is provided.}

Table \ref{tab:sobol-sim} shows the sample means and standard errors of the estimated Sobol indiceses for the four methods{, \color{black}where na\"{\i}ve PCE and SN-PCE use 500 and 100 observations, respectively}. Despite the vastly different sample sizes used for each method, the sample means are nearly identical across the methods (except for the non-influential inputs, which have total Sobol indices smaller than $10^{-1}$). This indicates the estimation bias is nearly zero for these methods. The standard errors are also very similar across the methods for the influential inputs, indicating that SN-PCE achieves a similar accuracy as the other methods with a much smaller sample size.   

% As shown in Table \ref{tab:sobol-sim2}, the SN-PCE method detects the non-influential inputs for network output by assigning small values or even $0$ to their first-order and total Sobol indices. More importantly, the SN-PCE achieves the similar accuracy level in estimating the first-order and total Sobol indices for the \textit{influential} inputs compared with the Monte Carlo method and the na\"{\i}ve PCE. However, the SN-PCE uses much fewer of observations than the other two methods. 
%The negative total Sobol indices calculated using the Monte Carlo method in Table \ref{tab:sobol-sim2} can be explained by their large standard errors since it is difficult for the Monte Carlo method to accurately estimate Sobol indices for non-influential inputs even with a large number of observations \cite{myshetskaya2008monte,owen2013better}.

\begin{figure}[H]
{\centering
\includegraphics[width=0.45\textwidth]{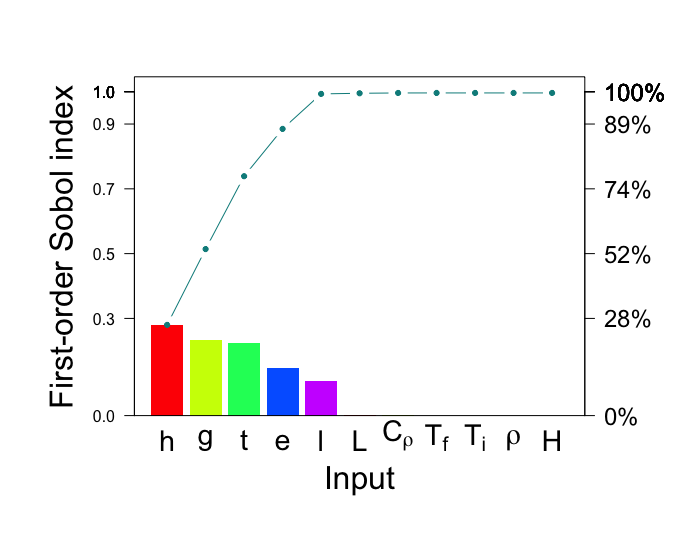} \\
\centering\caption{\small{Pareto chart of the first-order Sobol indices estimated using SN-PCE for the welding process. Higher bar indicates that the input has a more influence (excluding interactions with other inputs) on the variance of the process output $E$. }}
\label{fig:weld_process_pareto} 
}
\end{figure}

Fig.~\ref{fig:weld_process_pareto} presents a Pareto chart of the first-order Sobol indices estimated from SN-PCE. Along with Table \ref{tab:sobol-sim2}, the chart confirms that the weld zone dimensions ($h, g, t, e, l)$ are the most influential inputs for the variance of the process output $E$. Also, the cumulative sum of the first-order Sobol indices approaches 100\% in the chart, implying that the interactions between the inputs do not have significant effects on the variance of $E$. It echoes the finding from Table \ref{tab:sobol-sim2} that the first-order Sobol indices are approximately equal to the total Sobol indices across all the influential inputs. %And  goes approximately to $100\%$ in Fig.~\ref{fig:weld_process_pareto}. It suggests that the interaction terms among the inputs do not significantly affect the variance of output $E$. 

\begin{table*}[!ht]
\centering
\caption{Sample means and standard errors (rounded to two decimal places) of the estimated first-order and total Sobol indices based on 50 replications for the welding process. The inputs in the first column are sorted in descending order of the first-order Sobol indices estimated from the Monte Carlo method. For each replication,{\color{black} $\dagger$random sampling, $\ddagger$orthogonal array sampling, $\dagger\dagger$Na\"{\i}ve PCE, and $\ddagger\ddagger$SN-PCE use 10,000, 10,000, 500, and 100 observations, respectively.} For the influential inputs, SN-PCE attains similar standard errors as the other methods despite using the smallest sample size. } 
\label{tab:sobol-sim}
\begin{adjustbox}{width=0.85\textwidth}
\small
%\begin{tabular}{|c|c|c|c|c|c|c|c|c|c|}
\begin{tabular}{|c|c|c|c|>{\color{black}}c|>{\color{black}}c|c|c|c|c|}
\hline
\multirow{2}{*}{Input} & \multirow{2}{*}{Distribution} & \multicolumn{2}{c|}{$\text{Random Sampling}^{\dagger}$} &
\multicolumn{2}{>{\color{black}}c|}{$\text{Orthogonal Array}^{\ddagger}$}&
\multicolumn{2}{c|}{$\text{Na\"{\i}ve PCE}^{\dagger\dagger}$} & \multicolumn{2}{c|}{$\text{SN-PCE}^{\ddagger\ddagger}$} \\ \cline{3-10} 
 &  & \multicolumn{1}{c|}{First-order} & \multicolumn{1}{c|}{Total} &
\multicolumn{1}{>{\color{black}}c|}{First-order} &  \multicolumn{1}{>{\color{black}}c|}{Total}&
 \multicolumn{1}{c|}{First-order} &  \multicolumn{1}{c|}{Total} & \multicolumn{1}{c|}{First-order} & \multicolumn{1}{c|}{Total} \\ \hline
\multirow{2}{*}{$h$} & \multirow{2}{*}{$N(2.6, 0.5)$} & $2.76 \times 10^{-1}$ & $2.85 \times 10^{-1}$ \color{black} &$2.78 \times 10^{-1}$ & $2.81 \times 10^{-1}$ & $2.78 \times 10^{-1}$ & $2.80 \times 10^{-1}$ & $2.80 \times 10^{-1}$ & $2.81 \times 10^{-1}$ \\
 &  & $\pm 0.02 \times 10^{-1}$ & $\pm 0.04 \times 10^{-1}$ &  $\pm 0.01 \times 10^{-1}$ & $\pm 0.03 \times 10^{-1}$ &$\pm 0.01\times10^{-1}$ & $\pm 0.01 \times 10^{-1}$ & $\pm 0.02\times10^{-1}$ & $\pm 0.02\times10^{-1}$ \\
 \hline
 \cline{1-2}
\multirow{2}{*}{$g$} & \multirow{2}{*}{$N(2, 0.1)$} & $2.31 \times 10^{-1}$ & $2.30 \times 10^{-1}$ & $2.34 \times 10^{-1}$ & $2.32 \times 10^{-1}$ &$2.32 \times 10^{-1}$ & $2.33 \times 10^{-1}$ & $2.34 \times 10^{-1}$ & $2.35 \times 10^{-1}$ \\
 &  & $\pm 0.01\times 10^{-1}$ & $\pm 0.03 \times 10^{-1}$ & $\pm 0.01\times 10^{-1}$ & $\pm 0.03 \times 10^{-1}$ &$\pm 0.01\times10^{-1}$ & $\pm 0.01 \times 10^{-1}$ & $\pm 0.02\times10^{-1} $ & $ \pm 0.02\times10^{-1}$ \\
 \hline
 \cline{1-2}
\multirow{2}{*}{$t$} & \multirow{2}{*}{$N(15, 0.6)$} & $2.29 \times 10^{-1}$ & $2.24 \times 10^{-1}$ & $2.27 \times 10^{-1}$ & $2.32 \times 10^{-1}$ &$2.29 \times  10^{-1}$ & $2.29\times 10^{-1}$ & $2.25 \times 10^{-1}$ & $2.26 \times 10^{-1}$ \\
 &  & $\pm 0.01 \times 10^{-1}$ & $\pm 0.03 \times 10^{-1}$ & $\pm 0.01 \times 10^{-1}$ & $\pm 0.04 \times 10^{-1}$ & $\pm 0.01\times10^{-1}$ & $\pm 0.01 \times 10^{-1}$ & $\pm 0.02\times 10^{-1}$ & $\pm 0.02\times 10^{-1}$ \\ 
 \hline
 \cline{1-2}
\multirow{2}{*}{$e$} & \multirow{2}{*}{$N(11, 1)$} & $1.46 \times 10^{-1}$ & $1.48 \times 10^{-1}$ & $1.46 \times 10^{-1}$ & $1.50 \times 10^{-1}$ & $1.45\times 10^{-1}$ &$ 1.47 \times 10^{-1}$ &$1.46\times 10^{-1}$ & $1.48 \times 10^{-1}$ \\
 &  & $\pm 0.01\times10^{-1}$ & $\pm 0.04 \times 10^{-1}$ &$\pm 0.01\times10^{-1}$ & $\pm 0.04 \times 10^{-1}$ & $\pm 0.01\times10^{-1}$ & $\pm 0.01 \times 10^{-1}$ & $\pm 0.01\times 10^{-1}$ & $\pm 0.01\times 10^{-1}$ \\
\hline
\cline{1-2}
\multirow{2}{*}{$l$} & \multirow{2}{*}{$N(8.5, 0.5)$} & $1.07 \times 10^{-1}$ & $1.08 \times 10^{-1}$ &$1.08 \times 10^{-1}$ & $1.12 \times 10^{-1}$ & $1.07\times10^{-1}$ &$ 1.12\times10^{-1}$ & $1.08 \times 10^{-1}$ &$ 1.12 \times 10^{-1}$ \\
 & &$\pm 0.01 \times 10^{-1} $&$ \pm 0.04 \times 10^{-1} $& $\pm 0.01 \times 10^{-1} $&$ \pm 0.04 \times 10^{-1} $& $ \pm 0.00\times10^{-1} $&$ \pm 0.00\times10^{-1} $&$ \pm 0.01\times10^{-1} $&$ \pm 0.01\times10^{-1}$ \\
 \hline
 \cline{1-2}
\multirow{2}{*}{$L$} & \multirow{2}{*}{$N(500, 10)$} & $1.96 \times 10^{-3}$ & $4.45\times 10^{-4}$ & $1.95 \times 10^{-3}$ & $2.57 \times 10^{-3}$ & $1.96\times10^{-3}$ & $2.00\times10^{-3}$ & $1.94 \times 10^{-3}$ & $1.95 \times 10^{-3}$ \\
 &  & $\pm 0.01 \times 10^{-3}$ & $\pm 43.4 \times 10^{-4}$ & $\pm 0.01 \times 10^{-3}$ & $\pm 4.25 \times 10^{-3}$& $\pm 0.01\times10^{-3}$ & $\pm 0.01\times10^{-3}$ & $\pm 0.03\times10^{-3} $ & $\pm 0.03\times10^{-3}$ \\
 \hline
 \cline{1-2}
\multirow{2}{*}{$C_{\rho}$} & \multirow{2}{*}{$N(500, 5)$} & $9.67 \times 10^{-4}$ & $-9.76 \times 10^{-4}$ &$9.66 \times 10^{-4}$ & $4.94 \times 10^{-6}$ & $9.69\times10^{-4}$ & $9.89\times10^{-4}$ & $9.77 \times 10^{-4}$ & $9.77 \times 10^{-4}$ \\
 &  & $\pm 0.07\times10^{-4}$ & $\pm 43.02 \times 10^{-4}$ & $\pm 0.05 \times 10^{-4}$ & $\pm 0.04 \times 10^{-2}$ & $\pm 0.05\times10^{-4}$ & $\pm 0.05 \times 10^{-4}$ & $\pm 0.15\times10^{-4}$ & $\pm 0.15\times10^{-4}$ \\
 \hline
 \cline{1-2}
\multirow{2}{*}{$T_{f}$} & \multirow{2}{*}{$N(1628, 10)$} & $2.75 \times 10^{-4}$ & $-1.64 \times 10^{-3}$ & $2.76 \times 10^{-4}$ & $1.50 \times 10^{-6}$ & $2.75\times10^{-4}$ & $2.81\times10^{-4}$ & $7.55\times 10^{-5}$ & $7.55\times 10^{-5}$ \\
 &  & $\pm 0.02\times10^{-4}$ & $\pm 4.35 \times 10^{-3}$ & $\pm 0.05\times10^{-4}$ & $\pm 4.25\times 10^{-3}$ &$\pm 0.01\times10^{-4}$ & $\pm 0.01 \times 10^{-4}$ & $\pm 1.02\times10^{-5}$ & $\pm 1.02\times10^{-5}$ \\ 
 \hline
 \cline{1-2}
\multirow{2}{*}{$T_{i}$} & \multirow{2}{*}{$N(303, 0.3)$} & $8.32\times10^{-6}$ & $-1.74\times10^{-3}$ & $8.23\times10^{-6}$ & $2.76\times10^{-4}$ & $8.28\times10^{-6}$ & $8.45\times10^{-6}$ & \multirow{2}{*}{0} & \multirow{2}{*}{0} \\
 &  & $\pm 0.04\times10^{-6}$ & $\pm 0.04 \times10^{-3}$ &$\pm 0.05\times10^{-6}$ & $\pm 4.25 \times10^{-3}$ & $\pm 0.04\times10^{-6}$ & $\pm 0.04 \times 10^{-6}$ &  &  \\ 
 \hline
 \cline{1-2}
\multirow{2}{*}{$\rho$} & \multirow{2}{*}{$N(8238, 10)$} & $7.23 \times 10^{-6}$ & $-1.78 \times 10^{-3}$ & $7.20 \times 10^{-6}$ & $4.06 \times 10^{-8}$ & $7.16\times10^{-6}$ & $7.30\times10^{-6}$ & \multirow{2}{*}{0} & \multirow{2}{*}{0} \\
 &  & $\pm 0.05 \times 10^{-6}$ & $\pm 4.31 \times 10^{-3}$ & $\pm 4.06 \times 10^{-8}$ & $\pm 4.25 \times 10^{-3}$& $\pm 0.03\times10^{-6}$ & $\pm 0.03 \times 10^{-6}$ &  &  \\
 \hline
 \cline{1-2} \cline{7-8} 
\multirow{2}{*}{$H$} & \multirow{2}{*}{$N(2270, 3)$} & $3.32 \times10 ^{-10}$ & $-1.77 \times 10^{-3}$ &$3.30 \times10 ^{-10}$ & $1.94 \times 10^{-12}$ & $3.31\times10^{-10}$ & \multicolumn{1}{l|}{ $3.41\times10^{-10}$}& 
{\multirow{2}{*}{0}} & 
{\multirow{2}{*}{0}} \\
 &  & $\pm 0.02 \times 10^{-10}$ & $\pm 4.32 \times 10^{-3}$ & $\pm 0.02 \times 10^{-10}$ & $\pm 4.25 \times 10^{-3}$ & $\pm 4.25\times10^{-3}$ & \multicolumn{1}{l|}{$\pm 0.02\times10^{-10}$} & & \\ \cline{1-2} \cline{7-8} 
 \hline
\end{tabular}
 \end{adjustbox}
\end{table*}

%\hl{Ashis: So, how would these observations benefit anyone operating the welding process? Here, I think you need to emphasize the connection with quality/reliability.}
%%%%%%%%%%%%%%%%%%%%%%%%%%%%%%%%%%%%%%%%%%%%%%%%%%%%%%%%%%%%%%%%%%%%%%
\subsection{Injection molding process}
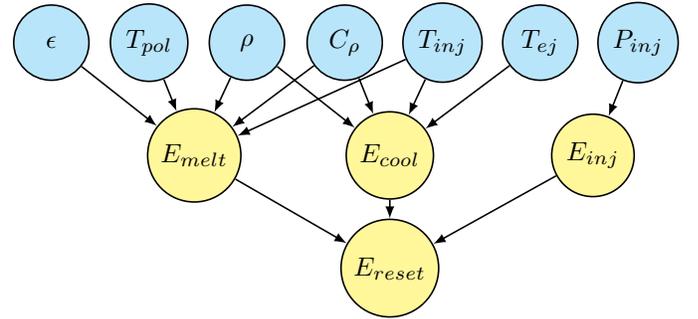
\begin{figure}[!ht]
{\centering
\begin {tikzpicture}[-latex ,auto ,node distance =1.5cm and 1.3cm ,on grid ,
semithick , state/.style ={ circle ,
draw, fill = cyan!25, text=black , minimum width =1 cm}, output/.style = {circle ,  draw, fill = yellow!50 , text=black , minimum width =1 cm}]
actor role/.style = {rectangle, draw=black!80, ultra thick,
    minimum size = 5mm, every actor role},
\node[state] (e) [xshift =0.1cm]{$\epsilon$ };
\node[state] (Tpol) [right =of e] {$T_{pol}$};
\node[state] (rho) [right =of Tpol] {$\rho$};
\node[state] (Cp) [right =of rho] {$C_{\rho}$};
\node[state] (Tinj) [right =of Cp] {$T_{inj}$};
\node[state] (Tej) [right =of Tinj] {$T_{ej}$};
\node[state] (Pinj) [right =of Tej] {$P_{inj}$};
\node[output] (Emelt) [below =of Tpol, xshift = 0.6cm] {$E_{melt}$};
\node[output] (Ecool) [below =of Cp, xshift = 0.6cm] {$E_{cool}$};
\node[output] (Einj) [below =of Pinj, xshift = -0.6cm] {$E_{inj}$};
\node[output] (Ereset) [below =of Ecool] {$E_{reset}$};
\path (e) edge [] node[below = 0.15 cm] {}(Emelt);
\path (Tpol) edge [] node[below = 0.15 cm] {}(Emelt);
\path (rho) edge [] node[below = 0.15 cm] {}(Emelt);
\path (Cp) edge [] node[below = 0.15 cm] {}(Emelt);
\path (Tinj) edge [] node[below = 0.15 cm] {}(Emelt);
\path (rho) edge [] node[below = 0.15 cm] {}(Ecool);
\path (Cp) edge [] node[below = 0.15 cm] {}(Ecool);
\path (Tinj) edge [] node[below = 0.15 cm] {}(Ecool);
\path (Tej) edge [] node[below = 0.15 cm] {}(Ecool);
\path (Pinj) edge [] node[below = 0.15 cm] {}(Einj);
\path (Emelt) edge [] node[below = 0.15 cm] {}(Ereset);
\path (Ecool) edge [] node[below = 0.15 cm] {}(Ereset);
\path (Einj) edge [] node[below = 0.15 cm] {}(Ereset);
\end{tikzpicture}
\centering\caption{\small{This DAG represents the relationships among the variables in the injection molding process \cite{Nannapaneni:2014}; the yellow shaded nodes $E_{melt}$, $E_{cool}$, and $E_{inj}$ are the energies consumed in three subprocesses that depend on different blue shaded network inputs. The network output $E_{reset}$ depends on $E_{melt}$, $E_{cool}$, and $E_{inj}$.   }}
\label{fig:inj_process}}
\end{figure}

The injection molding process has more intricate relationships between process variables than the welding process, as depicted in Fig.~\ref{fig:inj_process}. %in \cite{Nannapaneni:2014} is served as our second empirical application to validate the accuracy of the estimated Sobol indices using the SN-PCE. As it is shown in Fig.~\ref{fig:inj_process}, 
The process output of interest is the energy consumed in resetting the process, $E_{reset}$. This energy depends on the melting energy $E_{melt}$, the injection energy $E_{inj}$, and the cooling energy $E_{cool}$ as follows: $E_{reset} = 0.25\left(E_{melt} + E_{inj} + E_{cool}\right)$.
% \begin{equation}
% E_{reset} = \frac{1}{4}\left(E_{melt} + E_{inj} + E_{cool}\right).\nonumber
% \end{equation}
Here, %the melting energy $E_{melt}$ is modeled as
$E_{melt} = {P_{melt}\times V_{shot} }/{Q}$,
% \begin{equation}
% \begin{aligned}
% E_{melt} &= \frac{P_{melt}\times V_{shot} }{Q},
% \end{aligned}\nonumber
% \end{equation}
where
\begin{equation}
\begin{aligned}
P_{melt} &=\frac{1}{2}\left(\rho\times Q\times C_{p}\times \left(T_{inj}-T_{pol}\right) + \rho\times Q\times H_{f}\right), \\
V_{shot} &= V_{part}\times \left(1+\frac{\epsilon}{100}+\frac{\Delta}{100} \right).
\end{aligned}\nonumber
\end{equation}
Here, $\rho$ (specific gravity), $C_{p}$ (heat capacity), $T_{pol}$ (initial polymer temperature), $\epsilon$ (shrinkage parameter), and $T_{inj}$ (injection temperature) are the network inputs. $Q$ (flow rate), $H_{f}$ (polymer heat of fusion), $V_{part}$ (volume of mold), and $\Delta$ (buffer) are constant parameters. On the other hand, $E_{inj} = P_{inj} \times V_{part}$ and 
%The injection energy $E_{inj}$ and the cooling energy $E_{cool}$ are modeled separately as follows: 
\begin{equation}
\begin{aligned}
%E_{inj} &= P_{inj} \times V_{part}, \\
E_{cool} &= \frac{\rho\times V_{part}\times \left(C_{p}\times\left(T_{inj} -T_{ej}\right) \right)}{COP},
\end{aligned}\nonumber
\end{equation}
where $P_{inj}$ (injection pressure), $T_{ej}$ (ejection temperature), and $T_{inj}$ (injection temperature) are the network inputs. $COP$ (coefficient of performance of the cooling equipment) is a constant parameter. The network inputs' distributions are presented in Table \ref{tab:sobol-sim2}.

To adequately model the more intricate network structure of the injection molding process, we use the highest polynomial order of $4$ for both na\"{\i}ve PCE and SN-PCE, compared to $3$ used for the welding process; na\"{\i}ve PCE requires at least $330 =\binom{7+4}{4}$ observations for $\dim\!\left(\boldsymbol{\xi}\right)=7$ and network PCE (i.e., SN-PCE without sparsity) requires at least $126 =\binom{5+4}{4}$ observations because eq.~\eqref{eq:num_var_nPCE} equals 5. Thus, we use 500 and 200 observations for na\"{\i}ve PCE and SN-PCE, respectively, compared to 500 and 100 used for the welding process. Yet, again, SN-PCE could use even fewer observations because it identifies only $9$ orthonormal polynomials (3\% of those used in na\"{\i}ve PCE) as necessary to approximate the network output of this particular process across all 50 replications.

\begin{table}[H]
\centering
\caption{Sample means and standard errors (rounded to two decimal places) of the estimated first-order Sobol indices based on 50 replications for the injection molding process. For each replication, $\ddagger\ddagger$SN-PCE uses 200 observations. All the other setups are the same as in Table \ref{tab:sobol-sim}.} \label{tab:sobol-sim2}
\begin{adjustbox}{width=0.48\textwidth}
\small
%\begin{tabular}{|l|l|l|l|l|l|}
\begin{tabular}{|c|c|c|>{\color{black}}c|c|c|}
\hline
\multicolumn{1}{|c|}{Input} & \multicolumn{1}{c|}{Distribution} & \multicolumn{1}{c|}{$\text{Random Sampling}^{\dagger}$} &
\multicolumn{1}{>{\color{black}}c|}{$\text{Orthogonal Array}^{\ddagger}$}&
\multicolumn{1}{c|}{$\text{Naive PCE}^{\dagger\dagger}$} & \multicolumn{1}{c|}{$\text{SN-PCE}^{\ddagger\ddagger}$} \\ \hline
\multirow{2}{*}{$T_{inj}$} & \multirow{2}{*}{$N(210, 3)$} & $4.76\times 10^{-1}$ & $4.75\times 10^{-1}$& $4.77\times 10^{-1}$ & $4.78\times 10^{-1}$ \\
 &  & $\pm 0.02\times 10^{-1}$ &$\pm 0.02\times 10^{-1}$ & $\pm 0.01\times 10^{-1}$ & $\pm 0.02\times 10^{-1}$ \\
 \hline
 \cline{1-2}
\multirow{2}{*}{$T_{ej}$} & \multirow{2}{*}{$N(35, 3)$} & $2.61\times 10^{-1}$ & $2.61\times 10^{-1}$& $2.62\times 10^{-1}$ & $2.61\times 10^{-1}$ \\
 &  & $\pm 0.01\times 10^{-1}$ &$\pm 0.01\times 10^{-1}$ &$\pm 0.01\times 10^{-1}$ & $\pm 0.01\times 10^{-1}$ \\
 \hline
 \cline{1-2}
\multirow{2}{*}{$\rho$} & \multirow{2}{*}{$U(950, 990)$} & $2.27\times 10^{-1}$ & $2.27\times 10^{-1}$&$2.26\times 10^{-1}$ & $2.26\times 10^{-1}$ \\
 &  & $\pm 0.01\times 10^{-1}$ &$\pm 0.01\times 10^{-1}$ &$\pm 0.00\times 10^{-1}$ & $\pm 0.01\times 10^{-1}$ \\
 \hline
 \cline{1-2}
\multirow{2}{*}{$T_{pol}$} & \multirow{2}{*}{$N(40, 3)$} & $3.22\times 10^{-2}$ &$3.22\times 10^{-2}$ &$3.21\times 10^{-2}$ & $3.20\times 10^{-2}$ \\
 &  & $\pm 0.02\times 10^{-2}$ & $\pm 0.02\times 10^{-2}$&$\pm 0.02\times 10^{-2}$ & $\pm 0.02\times 10^{-2}$ \\
 \hline
 \cline{1-2}
\multirow{2}{*}{$C_{\rho}$} & \multirow{2}{*}{$U(2250, 2260)$} & $2.61\times 10^{-3}$ &$2.62\times 10^{-3}$& $2.61\times 10^{-3}$ & $2.61\times 10^{-3}$ \\
 &  & $\pm 0.01\times 10^{-3}$ &$\pm 0.01\times 10^{-3}$ &$\pm 0.01\times 10^{-3}$ & $\pm 0.01\times 10^{-3}$ \\
 \hline
 \cline{1-2}
\multirow{2}{*}{$\epsilon$} & \multirow{2}{*}{$U(0.018, 0.021)$} &  $7.76\times 10^{-9}$ & $7.74\times 10^{-9}$ &$7.76\times 10^{-9}$ & \multicolumn{1}{c|}{\multirow{2}{*}{$0$}} \\
 &  & $\pm 0.04\times 10^{-9}$ & $\pm 0.04\times 10^{-9}$&$\pm 0.03\times 10^{-9}$ & \multicolumn{1}{c|}{} \\
 \hline
 \cline{1-2}
\multirow{2}{*}{$P_{inj}$} & \multirow{2}{*}{$N(90, 4)$} & $4.75\times 10^{-14}$  &$4.73\times 10^{-14}$ & $4.77\times 10^{-14}$ & \multicolumn{1}{c|}{\multirow{2}{*}{$0$}} \\
 &  & $\pm 0.02\times 10^{-14}$ &$\pm 0.02\times 10^{-14}$ &$\pm 0.02\times 10^{-14}$ & \multicolumn{1}{c|}{} \\
 \hline
 \cline{1-2}
\end{tabular}
\end{adjustbox}
\end{table}

\begin{figure}[H]
{\centering
\includegraphics[width=0.45\textwidth]{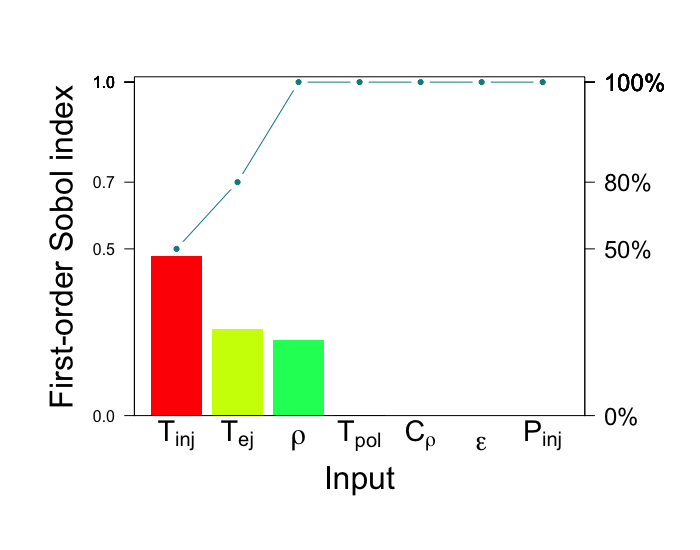}\\
\centering\caption{\small{Pareto chart of the first-order Sobol indices estimated using SN-PCE for the injection molding process. Higher bar indicates that the input has a more influence (excluding interactions with other inputs) on the variance of process output $E_{reset}$. }}
\label{fig:im_process} 
}
\end{figure}

Like the welding process, the injection molding process turns out to have the first-order Sobol indices approximately equal to the total Sobol indices for the influential inputs. Thus, we only report the first-order Sobol indices in Table \ref{tab:sobol-sim2}. Again, SN-PCE attains similar standard errors as the other methods for the influential inputs despite using fewer observations. Fig.~\ref{fig:im_process} shows that $T_{inj}$ determines nearly 50\% of the variance of network output $E_{reset}$. $T_{ej}$ and $\rho$ have comparable effects ($26\%$ and $23\%$, respectively), while other inputs, $T_{pol}$, $C_{\rho}$, $\epsilon$, and $P_{inj}$, barely influence the variance of $E_{reset}$.

\section{Conclusion}
\label{sec:conclusion}
This paper proposes SN-PCE to model uncertainty propagation in a broad class of processes represented as DAGs. %\textcolor{black}{Ashis: Model what? Uncertainty propagation? PCE is less useful as a process model per se.} 
A DAG encodes the dependencies between the variables in a process, including the process output (sink node in the DAG) and inputs (source nodes in the DAG). SN-PCE effectively captures how the inputs influence the output variance. %through the network-structured process. 
Theoretically, it is shown that network PCE (equivalent to SN-PCE without sparsity) is valid in the sense that its prediction of the output converges in probability to the true output under reasonable assumptions. Empirically, SN-PCE is shown to accurately estimate the Sobol indices of the output with respect to the inputs for two manufacturing processes {\color{black} and a flooding process}. SN-PCE uses substantially fewer observations than the black-box approaches ({\color{black}Monte Carlo sampling methods} and na\"{\i}ve PCE) to accurately identify the influential inputs, showing promise for efficient sensitivity analysis in process automation. 

%to identify the influential inputs in explaining the output variance of an engineering system which can be formed as a DAG. Comparing with the existing methods, the SN-PCE requires much fewer number of observations. The proposed method is validated through both theoretical analysis and empirical results. On one hand, the theoretical analysis proves that under certain regularity conditions, the model prediction of the output converges in probability to the true output. On the other hand, two manufacturing applications, namely, the welding process and the injection molding process, are used to empirically validate the accuracy of obtaining the Sobol indices using our proposed model comparing with a state-of-the-art Monte Carlo-based approach. The empirical results show that the SN-PCE can achieve the same level of accuracy as the Monte Carlo approach using fewer number of observations. 

Future research may investigate extension of the proposed model to even more general networks. One direction could be to relax the assumption of mutual independence of the network inputs. While their dependencies would complicate the estimation and interpretation of the sensitivity indices, the indices proposed in \cite{liu2018data} might help decode how the dependent network inputs influence the output. Another research direction would be to allow cycles (or, feedback loops) in the network, thereby, extending this work beyond directed acyclic networks. Lastly, while domain experts can often identify DAGs of their systems, machine learning can more efficiently identify complex DAGs from data under the causal Markov and causal faithfulness conditions \cite{meganck2006learning,huang2013sparse}. Future research may investigate the best way to combine network identification methods with the proposed method in this paper.  %A new model may treat the nodes forming the cycles as a single node representing a random vector on a different network  
\section*{Acknowledgment}
This work was partially supported by the National Science Foundation (NSF grant CMMI-1824681). {\color{black}We thank the Editor, Associate Editor, and three anonymous reviewers for their feedback that helped significantly improve this paper.} 
%\addtolength{\textheight}{-12cm}  
%\clearpage
%\addtolength{\textheight}{-12cm}

\bibliographystyle{IEEEtran} %apalike
%\balance

\bibliography{myref.bib}
\vspace{-0.6cm}
%\balance
\begin{IEEEbiography}[{\includegraphics[width=1in,height=21in,clip,keepaspectratio]{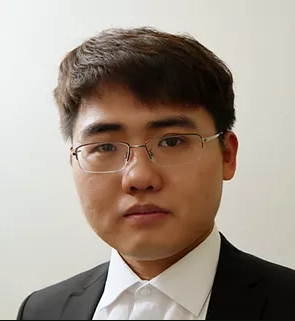}}]{Zhanlin Liu}
received the B.S. degree in Statistics from University of Iowa, Iowa City, IA, USA, in 2014, and the M.S. degree in Statistics from the University of Washington, Seattle, WA, USA, in 2016. 

He is currently working toward the Ph.D. degree in Industrial $\&$ Systems Engineering at the University of Washington, Seattle, WA, USA. 
His current research interests include uncertainty quantification, reliability analysis, and statistical modeling. 
\end{IEEEbiography}
\vspace{-0.6cm}
\begin{IEEEbiography}[{\includegraphics[width=1in,height=2in,clip,keepaspectratio]{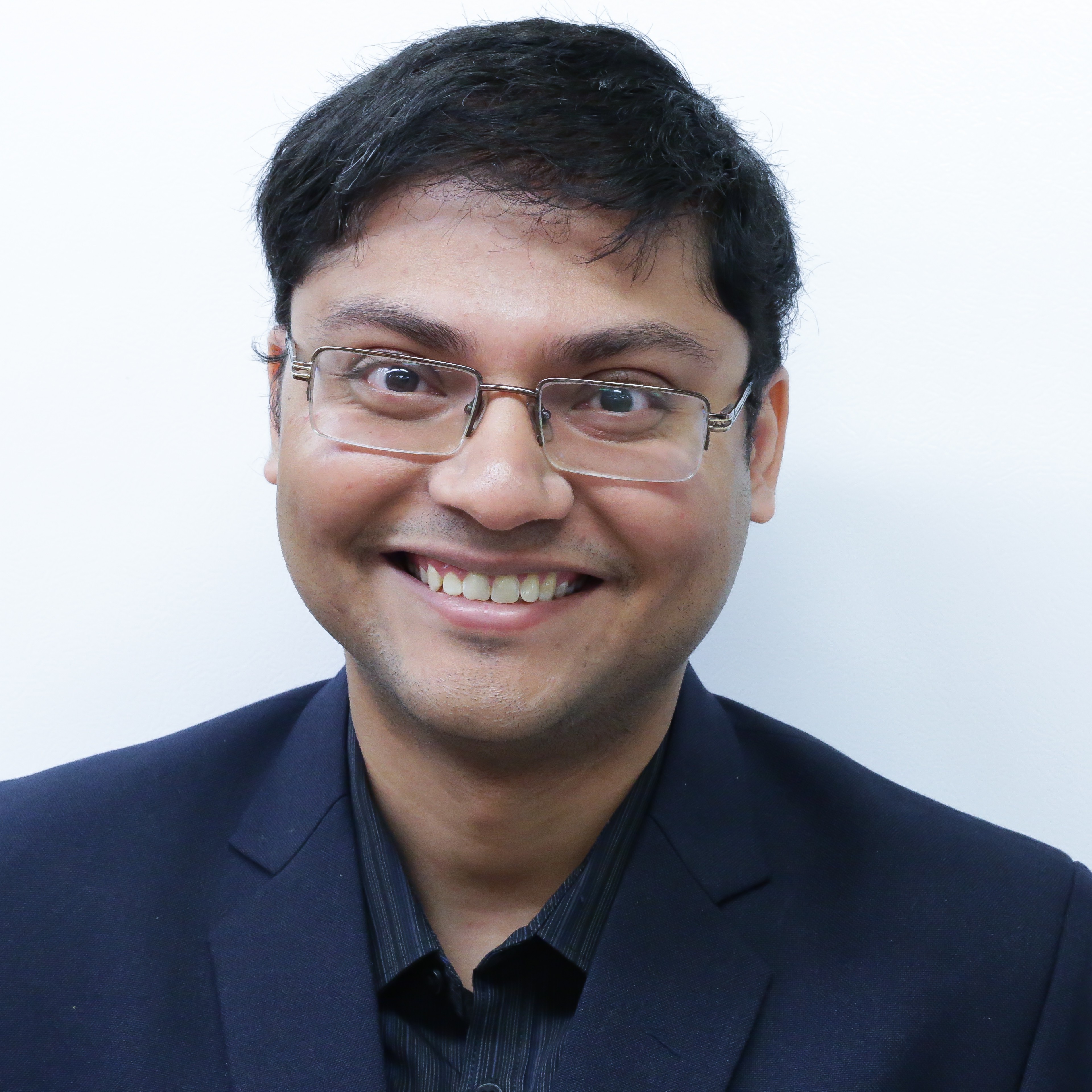}}]{Ashis G. Banerjee} (S'08-M'09) received the B.Tech. degree in Manufacturing Science and Engineering from the Indian Institute of Technology Kharagpur, Kharagpur, India, in 2004, the M.S. degree in Mechanical Engineering from the University of Maryland (UMD), College Park, MD, USA, in 2006, and the Ph.D. degree in Mechanical Engineering from UMD in 2009.

He is currently an Assistant Professor of Industrial \& Systems Engineering and Mechanical Engineering at the University of Washington, Seattle, WA, USA. Prior to this appointment, he was a Research Scientist at GE Global Research, Niskayuna, NY, USA. Previously, he was a Research Scientist and Postdoctoral Associate in the Computer Science and Artificial Intelligence Laboratory at Massachusetts Institute of Technology, Cambridge, MA, USA. His research interests include digital manufacturing, predictive and prescriptive analytics, and autonomous robotics.
\end{IEEEbiography}

\vspace{-0.6cm}
\begin{IEEEbiography}[{\includegraphics[width=1in,height=2in,clip,keepaspectratio]{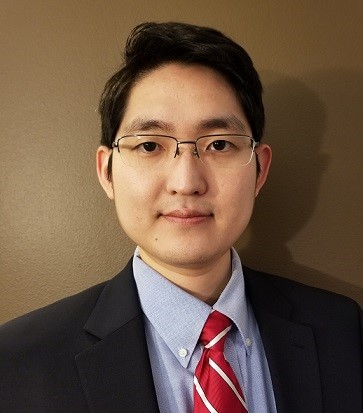}}]{Youngjun Choe} 
received the B.Sc. degrees in Physics and Management Science from KAIST, Korea in 2010, and both M.A. in Statistics and  Ph.D. in Industrial \& Operations Engineering from the University of Michigan, Ann Arbor, MI, USA in 2016. 

He is currently an Assistant Professor of Industrial \& Systems Engineering at the University of Washington, Seattle, WA, USA. His research centers around developing statistical methods to infer on extreme events. % using empirical and simulated data. 
\end{IEEEbiography}
%Youngjun, I think your bio is not in the IEEE recommended format. %Ashis, I updated the bio using your format. Thanks!

\end{document}